\documentclass[letterpaper,twocolumn,10pt]{article}
\usepackage{microtype}
\usepackage{graphicx}
\usepackage{subcaption}
\usepackage{booktabs}
\usepackage{multirow}
\usepackage{array}
\usepackage{hegroup}
\usepackage{makecell}
\usepackage{xspace}
\usepackage{enumitem}
\usepackage{algorithm}
\usepackage{algorithmic}
\newcommand{\mypara}[1]{\smallskip\noindent{\bf {#1}.}\xspace}
\usepackage{hyperref}
\usepackage{amsmath}
\usepackage{amssymb}
\usepackage{mathtools}
\usepackage{amsthm}
\usepackage[capitalize,noabbrev]{cleveref}
\theoremstyle{plain}

\theoremstyle{definition}

\theoremstyle{remark}

\usepackage[textsize=tiny]{todonotes}
\newcommand{\method}{$\mathsf{MARS}$\xspace}
\begin{document}

\title{Escaping the Linearity Trap: Manifold Detours for Black-Box Adversarial Attacks on Singing Audio Deepfake Detection}
\date{}

\author{
Yifan Liao\textsuperscript{1,2}\thanks{Equal contribution.} \ \ \
Yule Liu\textsuperscript{1,2}\footnotemark[1] \ \ \
Zhen Sun\textsuperscript{2} \ \ \
Zongmin Zhang\textsuperscript{2} \ \ \\
Yupeng He\textsuperscript{2} \ \ \
Jiaheng Wei\textsuperscript{2} \ \ \
Xinhu Zheng\textsuperscript{2} \ \ \
Xinlei He\textsuperscript{1}\thanks{Corresponding author: Xinlei He (\href{mailto:xinlei.he@whu.edu.cn}{xinlei.he@whu.edu.cn})} \ \ \
\\
\\
\textsuperscript{1}\textit{Wuhan University} \ \ \\
\textsuperscript{2}\textit{The Hong Kong University of Science and Technology (Guangzhou)} \ \ \\
}

\maketitle

\begin{abstract}
Recent advances in Singing Voice Synthesis (SVS) enable highly realistic singing voices, but also facilitate malicious uses such as unauthorized AI covers.
To mitigate these risks, singing voice deepfake detections (SVDD) have been developed as a primary technical defense.
Among them, Self-Supervised Learning (SSL)-based detectors have achieved state-of-the-art performance by fine-tuning speech SSL backbone together with the detection head to adapt to the singing domain and make them sensitive to singing-specific spoof artifacts, showing strong detection performance.
To bypass detection, existing adversarial attacks usually craft adversarial perturbations on the audio but often perform poorly in SSL-based SVDD.
This may create a misleading impression that SSL-based SVDD are inherently robust to adversarial perturbations.
However, we find that such apparent robustness largely stems from two unresolved challenges in existing attacking methods.
Firstly, at the \emph{optimization objective level}, conventional attacks optimize cross-entropy on local surrogate detectors, making perturbations prone to crossing surrogate-specific boundaries rather than suppressing spoof evidence shared by unseen detectors.
Secondly, at the \emph{optimization method level}, transfer-enhanced attacks stabilize or diversify gradient updates, but still follow the surrogate model's dominant direction.
In SSL-based SVDD, this direction can align with artifact-sensitive directions learned during detector fine-tuning, reducing its transferability to unseen detectors.
We refer to this geometric failure mode as the \textit{Linearity Trap}.

To better evaluate SVDD robustness, we propose \method (\underline{M}eta-\underline{A}dversarial \underline{R}egression of \underline{S}emantics), a transfer-based black-box robustness assessment framework tailored to SSL-based SVDD.
At the optimization objective level, \method shifts from surrogate-boundary crossing to hypothesis-evidence manipulation by constructing a natural semantic anchor from the pre-trained SSL space and an artifact anchor from the fine-tuned surrogate SSL space.
At the optimization method level, \method escapes the \textit{Linearity Trap} through a bi-level optimization strategy: the inner stage induces tangential exploration away from the direct path, while the outer stage guide the perturbed audio toward the natural semantic manifold.
Extensive experiments on the CtrSVDD benchmark show that \method substantially improves Attack Success Rate (ASR) under in-distribution transfer ($13\%$), out-of-distribution transfer ($10\%$), and cross-task evaluation ($36\%$), highlighting the urgent need to improve the robustness of the SVDD systems.
\end{abstract}

\section{Introduction}
Recent advances in Singing Voice Synthesis (SVS) has made it increasingly easy to synthesize realistic singing voices with convincing timbre, pronunciation, pitch contours, and musical expressiveness~\cite{DBLP:conf/aaai/Liu00CZ22,yamamoto2023nnsvs,shi2022muskits}.
This technique enables creative applications such as virtual singers.
However, they also lower the barrier for malicious misuse.
For instance, attackers can generate counterfeit singing voices that imitate real singers, produce unauthorized AI covers.
Such incidents have already raised public concerns about copyright infringement, identity theft, and the authenticity of online musical content~\cite{bbc2023drakeai}.

To mitigate these risks, singing voice deepfake detections (SVDD) have become a primary technical defense~\cite{almutairi2022review,amerini2025deepfake}.
Mainstream high-performing SVDD systems typically follow a self-supervised learning-based(SSL) detection paradigm~\cite{zhang2024svdd,zang2024singfake,zhang2024xwsb}: a pre-trained speech SSL model is used as the acoustic backbone, and a detection head is trained to classify bonafide and synthetic singing audio.
During training, the speech SSL backbone is often fine-tuned together with the detection head, adapting speech-pretrained representations to the singing domain and making them sensitive to singing-specific spoof artifacts.
This paradigm has achieved strong detection performance on SVS deepfake benchmarks.

To bypass deepfake detection, existing adversarial attacks ~\cite{zhang2020black,farooq2025transferable,lin2019nesterov,wang2021variance,long2025negative,xie2019improving,dong2019evading,wu2021adversarial,chen2025awt} usually craft adversarial perturbations on the audio but often show limited performances when applied to SSL-based SVDD.
The weak attack performance may create a misleading impression that SSL-based SVDD systems are inherently robust to adversarial attacks.
However, we argue that such apparent robustness mainly comes from the mismatch between existing attack designs and the representation structure of the SSL-based SVDD.
This mismatch becomes especially critical in realistic transfer-based black-box deployments.
SVDD systems are usually proprietary: the detection architectures are hidden from attackers.
Therefore, an adversary can only use public SSL models and local surrogate detectors to craft adversarial samples offline and transfer it to unknown victim detectors.
This setting poses two unresolved challenges in existing attack methods.

\mypara{Challenge 1: Optimization objective leads to overfitting to local surrogate}
At the optimization objective level, existing attacks remain centered on surrogate-boundary crossing.
Conventional audio attacks usually optimize cross-entropy on local surrogate detectors~\cite{zhang2020black,farooq2025transferable}.
Such objectives can make adversarial examples cross a particular surrogate decision boundary, but they do not explicitly suppress spoof evidence shared by unseen detectors.
This is especially problematic for SSL-based SVDD because different detectors may use different SSL backbones, aggregation modules and detection heads.
Therefore, a perturbation that succeeds on one surrogate detector may only move the sample across a model-specific boundary without placing it into a general bonafide representation region shared by unseen SSL-based detectors.

\mypara{Challenge 2: Optimization method leads to trajectory bias}
At the optimization method level, existing transfer-enhanced attacks improve black-box transfer by stabilizing or diversifying gradient updates through different methods~\cite{lin2019nesterov,wang2021variance,long2025negative,xie2019improving,dong2019evading,wu2021adversarial,chen2025awt}.
These methods make the attack follow surrogate gradients more reliably, but they still optimize the perturbation locally on the surrogate detector.
Therefore, the perturbation tends to move along the surrogate's locally dominant direction, rather than exploring detector-invariant directions in SSL feature space.
In SSL-based SVDD, this locally optimal direction may coincide with the artifact-sensitive direction learned during detector fine-tuning.
As a result, the perturbation remains coupled with the surrogate representation geometry, making it difficult to induce a detector-invariant shift and reducing its transferability to unseen detectors.
We refer to this failure mode as the \textit{Linearity Trap}.

To better evaluate the robustness of SVDD, we propose \method (\underline{M}eta-\underline{A}dversarial \underline{R}egression of \underline{S}emantics), a transfer-based black-box security assessment framework tailored to SSL-based SVDD.
\method addresses the above challenges at both the optimization objective level and the optimization method level.
At the optimization objective level, \method shifts from surrogate-boundary crossing to hypothesis-evidence manipulation.
Following the Neyman-Pearson lemma~\cite{neyman1933ix}, we model SVDD as a likelihood-ratio test between the bonafide hypothesis $H_0$ and the spoof hypothesis $H_1$.
This view suggests that an effective attack should reduce the evidence supporting $H_1$ while preserving or increasing the evidence aligned with $H_0$.
To instantiate this idea in SSL-based SVDD, \method constructs a natural semantic anchor from the pre-trained SSL space and an artifact anchor from the fine-tuned detector space.
Based on these two anchors, \method further performs Push-Pull steering: it pulls the adversarial example toward the natural semantic anchor while pushing it away from the spoof-sensitive artifact anchor.
This optimization objective encourages the adversarial example to preserve natural singing semantics while reducing the artifact evidence used by fine-tuned SVDD detectors.
At the optimization method level, \method escapes the \textit{Linearity Trap} through a bi-level optimization strategy.
The inner stage induces tangential exploration away from the direct path, forcing the perturbation to move off the dominant surrogate direction.
The outer stage then applies Push-Pull steering to guide the perturbed audio toward the natural semantic manifold and away from spoof-sensitive artifact evidence.
This curvilinear trajectory allows adversarial examples to avoid over-reliance on surrogate-specific directions and improves transferability to unseen SSL-based detectors.

We conduct extensive experiments demonstrating that \method substantially improves the Attack Success Rate (ASR) under in-distribution transfer (13\%), out-of-distribution transfer (10\%), and cross-task evaluation (36\%).
Comprehensive ablation studies further validate the effectiveness of our geometry-aware modeling and the proposed bi-level optimization strategy.
Additional analyses on stealthiness and robustness show that \method preserves high perceptual quality while remaining effective under defenses.

Overall, we make the following contributions: 
\begin{itemize}[leftmargin=10pt] 
    \item We conduct a geometric analysis of adversarial transferability within SSL foundation models, uncovering the Linearity Trap where traditional attacks fail by traversing the predictable geodesic path. 
    \item We propose \method, a meta-adversarial framework inspired by hypothesis testing and instantiated as a local geometric surrogate. By modeling the SSL feature space with von Mises-Fisher distributions and employing a bi-level Push-Pull optimization, \method generates perturbations that effectively evade detectors. 
    \item Extensive experiments demonstrate that \method achieves superior attack potency outperforming state-of-the-art attacks by over 13\% in average ASR against unknown black-box detectors (e.g., AASIST2, SLS, MultiConv), revealing severe vulnerabilities in current defenses. 
\end{itemize}

\section{Background and Related Work}
\mypara{Singing Voice Deepfake Detection (SVDD)}
Beyond SVS (singing voice synthesis), singing voice conversion (SVC) has become another major source of high-fidelity forged vocals.
Recent SVC systems include diffusion-based generation for high naturalness~\cite{DBLP:conf/asru/LiuCSM21}, one-shot conversion with only a short reference~\cite{DBLP:conf/interspeech/LiLS22}, and waveform-GAN variants improved by harmonic conditioning for better pitch/timbre stability~\cite{DBLP:conf/icassp/GuoZML22}.
Standardized evaluations such as the Singing Voice Conversion Challenge further accelerate SVC progress and broaden the attack surface faced by SVDD~\cite{DBLP:journals/corr/abs-2306-14422,DBLP:journals/corr/abs-2509-15629}.
SVDD has thus emerged as a dedicated detection task, where melody-driven prosody and accompaniment can mask synthesis artifacts.
SingFake~\cite{DBLP:conf/icassp/Zang0HD24} introduces an in-the-wild benchmark and shows that speech-oriented countermeasures degrade on singing data.
CtrSVDD~\cite{DBLP:conf/interspeech/ZangS0YHTXZGTD24} provides a controlled benchmark for systematic generalization analysis, and the SVDD Challenge~\cite{DBLP:conf/slt/ZhangZSYTD24} consolidates evaluation settings.
Additionally, domain-aware modeling has been explored; e.g., SingGraph~\cite{DBLP:conf/interspeech/ChenWJL24} incorporates lyric cues for singing-specific robustness. However, according to the recent report~\cite{DBLP:conf/slt/ZhangZSYTD24}, high-performing SVDD systems increasingly adopt an SSL-based detection paradigm.

These SSL models are pre-trained on large-scale unlabeled audio using objectives such as contrastive prediction to enable them to learn general acoustic representations that capture phonetic content, speaker traits, prosody, and other high-level speech cues.
Because of this strong transferability, SSL backbones such as Wav2Vec 2.0, HuBERT~\cite{hsu2021hubert}, WavLM~\cite{chen2022wavlm}, XLS-R~\cite{baevski2020wav2vec,babu2021xls}, and UniSpeech~\cite{chen2022unispeech} have become widely used in downstream speech tasks, including ASR, speaker verification, and spoofing detection. ~\cite{tak2022automatic,chen2022large}

\begin{figure}[t]
    \centering
    \includegraphics[width=0.95\linewidth]{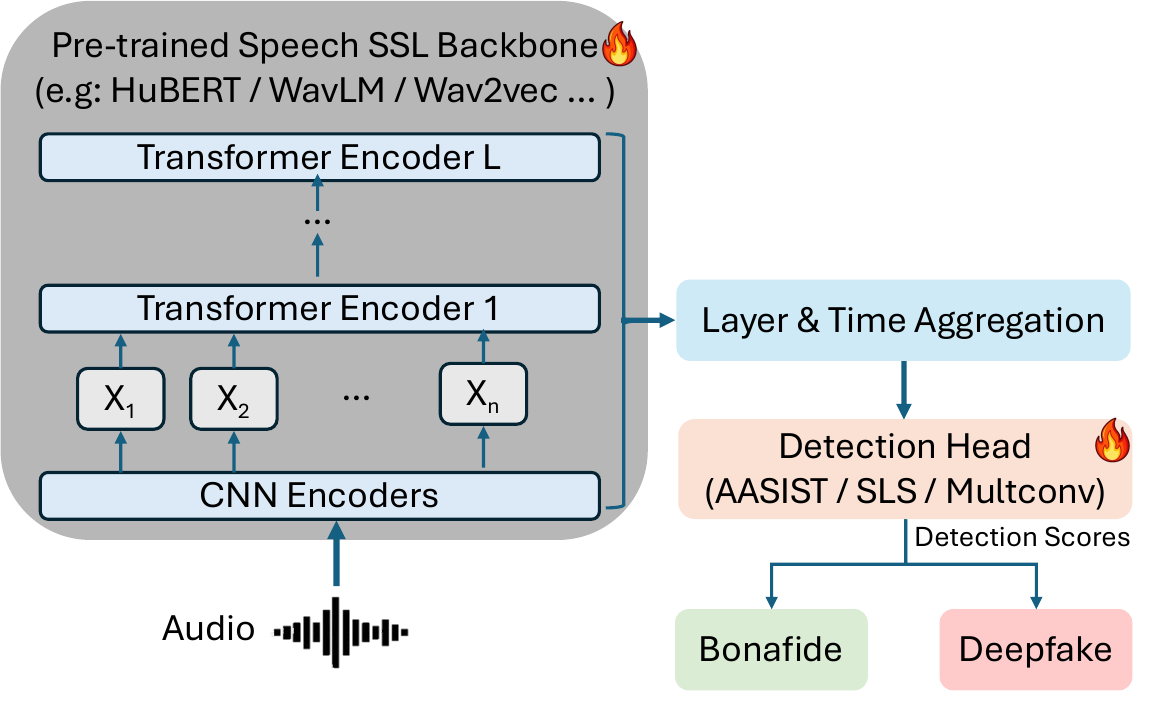}
    \caption{SSL-based SVDD framework. A pre-trained speech SSL backbone is adapted to the singing domain and jointly fine-tuned with a task-specific detection head for the classification.}
    \label{fig:SSL}
\end{figure}

However, to apply speech SSL models to singing voice deepfake detection, the models must be adapted to the singing domain.
As shown in~\Cref{fig:SSL}, mainstream high-performing SVDD systems typically follow an SSL-based detection framework.
Given an input singing audio $x$, a pre-trained speech SSL backbone first extracts contextual frame-level representations from multiple Transformer layers.
These layer-wise representations are then aggregated across layers and time by an aggregation module, and the resulting features are fed into a task-specific detection head to predict the bonafide/spoof label.

During training, the SSL backbone is adapted from speech to singing audio and jointly fine-tuned with the detection head on singing deepfake datasets.
This joint optimization allows the backbone to capture singing-domain characteristics, such as melody-driven prosody, sustained vowels, vibrato, expressive timbre variation, and accompaniment interference, while the detection head learns spoof-sensitive evidence for bonafide/deepfake classification.

\mypara{Adversarial Attacks on Deepfake Detection}
Gradient-based adversarial examples can actively evade deepfake and spoofing detectors in both white-box and transfer-based black-box settings.
FGSM performs a single-step sign-gradient update~\cite{DBLP:journals/corr/GoodfellowSS14}, while PGD iterates FGSM-style updates with projection (e.g., $\ell_\infty$ constraints) for stronger attacks~\cite{DBLP:conf/iclr/MadryMSTV18}.
Such attacks have been shown to be effective against spoofing countermeasures~\cite{DBLP:conf/asru/LiuWLM19}, and psychoacoustic masking can further enforce imperceptibility in audio~\cite{DBLP:conf/icml/QinCCGR19}.
However, adversarial robustness for SVDD remains underexplored, especially under transfer-based black-box settings with constraints that must preserve singing quality and musical pitch/rhythm structure.
Motivated by this gap, we compare against \emph{Ensemble-PGD} and a naive \emph{Joint Optimization} baseline, and develop a dedicated SVDD attack with perceptual and music-structure-preserving constraints. Beyond audio-specific attacks, the image domain has developed a rich line of transferable adversarial attacks for black-box evaluation.
Nesterov accelerated and scale-invariant attacks stabilize update directions with look-ahead gradients and multi-scale inputs~\cite{lin2019nesterov}.
Variance tuning reduces local gradient fluctuations by estimating gradient variance around the current point~\cite{wang2021variance}.
Negative Hessian trace regularization improves transferability by avoiding sharp surrogate-specific loss regions~\cite{long2025negative}.
Input diversity applies random transformations, such as resizing and padding, to reduce overfitting to a single input view~\cite{xie2019improving}.
Translation-invariant attacks smooth gradients with convolutional kernels to improve robustness under spatial shifts~\cite{dong2019evading}.
Adversarial transformations optimize perturbations to remain effective under transformation-induced variations~\cite{wu2021adversarial}.
Adversarial weight tuning perturbs surrogate weights during attack generation to reduce dependence on one fixed model parameterization~\cite{chen2025awt}.
Although effective in vision tasks, these methods do not directly address the representation geometry of SSL-based audio detectors.
Image-domain transfer attacks mainly improve perturbation robustness to spatial transformations, local gradient variance, or surrogate parameter changes.
However, in singing audio deepfake detection, spoof evidence is not a local spatial pattern; it is entangled with temporal dynamics, harmonic structure, pitch trajectories, timbre, and synthesis artifacts distributed across SSL layers.
Consequently, stabilizing, smoothing, diversifying, or reweighting surrogate gradients may still push the adversarial example along the detector's dominant artifact-sensitive direction, which unseen detectors are well positioned to reject.
The key challenge is therefore not merely to obtain a better surrogate gradient, but to manipulate the bonafide/spoof evidence encoded in SSL feature space.
Our method addresses this by explicitly modeling the two hypotheses in SSL representation space and using a bi-level detour mechanism to escape the dominant artifact-sensitive direction before steering toward the natural semantic manifold.

\section{Preliminaries}
\label{sec:preliminaries}

\mypara{Notations}
Let $\mathbf{x}\in\mathbb{R}^T$ be an input audio and
$\mathbf{x}_{\mathrm{adv}}=\mathbf{x}+\boldsymbol{\delta}$ be its adversarial version.
For an encoder $f$, we denote its normalized representation as
$\mathbf z_f(\mathbf x)=f(\mathbf x)/\|f(\mathbf x)\|_2$.
We use $\mathbf z_{\mathrm{base}}(\cdot)$ and
$\mathbf z_{\mathrm{det}}(\cdot)$ to denote normalized representations extracted
by the frozen foundation encoder and the fine-tuned detector encoder,
respectively.
These two representations are used to construct two local evidence terms.
We do not assume that they define a single shared global likelihood model.

To instantiate an evidence-ratio view in SSL representation space, we define two instance-wise anchors from the original input $\mathbf{x}$:
\begin{itemize}[leftmargin=*]
    \item \textit{Natural Anchor ($\mathbf{z}_0$):}
    $\mathbf{z}_0 \triangleq \mathbf{z}_{\mathrm{base}}(\mathbf{x})$,
    which serves as a local proxy for bonafide-related semantic and acoustic evidence.
    \item \textit{Artifact Anchor ($\mathbf{z}_1$):}
    $\mathbf{z}_1 \triangleq \mathbf{z}_{\mathrm{det}}(\mathbf{x})$,
    which serves as a local proxy for spoof-related artifact evidence.
\end{itemize}

\mypara{Evidence-Ratio Motivation}
For two simple hypotheses with known class-conditional densities, the Neyman-Pearson lemma states that thresholding the likelihood ratio gives the most powerful test at a fixed false-alarm rate~\cite{neyman1933ix}.
This classical result motivates an evidence-ratio perspective: a detector may be viewed as comparing evidence for the spoof hypothesis against evidence for the bonafide hypothesis.
However, in our transfer-based black-box SVDD setting, the true victim densities, decision threshold, logits, and internal representations are unavailable.
Therefore, MARS does not optimize the true Neyman-Pearson likelihood ratio and does not claim NP-optimality against unknown victim detectors.
Instead, we use this perspective to construct a local representation-level evidence surrogate from the two anchors above.

\mypara{Von Mises-Fisher Approximation}
Many SSL audio encoders use contrastive or normalized representation learning objectives, making angular similarity a natural measure in their feature spaces~\cite{wang2020understanding}.
A standard distribution for directional data on the unit hypersphere $\mathbb S^{d-1}$ is the von Mises-Fisher (vMF) distribution~\cite{de2020power}.
Given a mean direction $\boldsymbol{\mu}$ with $\|\boldsymbol{\mu}\|_2=1$ and a
concentration parameter $\kappa\ge 0$, its density is
\begin{equation}
    f_{\mathrm{vMF}}(\mathbf z;\boldsymbol{\mu},\kappa)
    =
    C_d(\kappa)\exp(\kappa \boldsymbol{\mu}^{\top}\mathbf z),
    \label{eq:vmf}
\end{equation}
where $C_d(\kappa)$ is a normalization constant. Thus, up to constants,
\begin{equation}
    -\log f_{\mathrm{vMF}}(\mathbf z;\boldsymbol{\mu},\kappa)
    \doteq
    -\kappa \boldsymbol{\mu}^{\top}\mathbf z.
    \label{eq:vmf_nll}
\end{equation}
This property allows us to convert local evidence models into cosine-similarity objectives.

\mypara{Local Evidence Surrogate}
We model bonafide-related semantic evidence and spoof-related artifact evidence with two local vMF models centered at the instance-wise anchors:
\begin{align}
q_0(\mathbf z_{\mathrm{base}}(\mathbf x_{\mathrm{adv}})\mid \mathbf x)
&\propto
\exp\left(
\kappa_0
\mathbf z_{\mathrm{base}}(\mathbf x_{\mathrm{adv}})^\top
\mathbf z_0
\right), \\
q_1(\mathbf z_{\mathrm{det}}(\mathbf x_{\mathrm{adv}})\mid \mathbf x)
&\propto
\exp\left(
\kappa_1
\mathbf z_{\mathrm{det}}(\mathbf x_{\mathrm{adv}})^\top
\mathbf z_1
\right).
\end{align}
The resulting local evidence score is
\begin{equation}
L_{\mathrm{evid}}(\mathbf x_{\mathrm{adv}};\mathbf x)
=
\log q_1(\mathbf z_{\mathrm{det}}(\mathbf x_{\mathrm{adv}})\mid \mathbf x)
-
\log q_0(\mathbf z_{\mathrm{base}}(\mathbf x_{\mathrm{adv}})\mid \mathbf x).
\end{equation}
Substituting the vMF forms gives, up to constants,
\begin{equation}
L_{\mathrm{evid}}
\doteq
-\kappa_0
\mathbf z_{\mathrm{base}}(\mathbf x_{\mathrm{adv}})^\top \mathbf z_0
+
\kappa_1
\mathbf z_{\mathrm{det}}(\mathbf x_{\mathrm{adv}})^\top \mathbf z_1.
\label{eq:local_evidence}
\end{equation}
Minimizing Eq.~\eqref{eq:local_evidence} therefore pulls the adversarial example toward the natural semantic anchor while pushing it away from the artifact anchor. 
Importantly, this is a local surrogate objective, not the exact likelihood ratio of the victim detector. We empirically validate that this anchor-based score correlates with surrogate detector evidence and report additional angular-model diagnostics in \Cref{app:anchor_diagnostic}.

\section{Threat Model}
\label{sec:threat_model}

As singing voice synthesis and conversion systems become increasingly realistic, singing deepfakes can be used to imitate specific singers or generate unauthorized AI covers.
SVDD is therefore becoming an important technical defense for copyright holders and forensic services.
However, if an adversary can slightly modify a deepfake song so that it is still perceived as the same singing content by human listeners but is classified as bonafide by an SVDD system, the detector can be bypassed before the content is uploaded, shared, or monetized.

\mypara{White-box and Black-box Settings}
A white-box adversary has full access to the victim detector, including its architecture, parameters, gradients, logits, feature representations, and training configuration.
Under this setting, the adversary can directly optimize the perturbation against the victim loss, making the attack considerably easier.
However, this assumption is unrealistic for deployed SVDD systems, which are typically proprietary and expose neither internal model details nor gradients.

In contrast, a black-box adversary has no access to the victim detector's internal information.
Depending on whether the attacker can interact with the victim system, black-box attacks can be further divided into query-based and transfer-based settings.
In a query-based setting, the attacker can repeatedly submit audio samples and use the returned labels or scores to estimate the decision boundary.
In a transfer-based setting, the attacker does not rely on victim queries during optimization.
Instead, adversarial examples are generated on local surrogate models and then transferred to unknown victim detectors.
We focus on this transfer-based black-box setting because it is more restrictive and better reflects real-world SVDD deployment, where platforms may hide confidence scores and have rate-limits.

\mypara{Adversary's Goal}
Let $\mathbf{x} \in \mathbb{R}^T$ be a deepfake singing audio sample that is correctly detected as spoof by a victim detector $f_v$.
The adversary aims to generate an adversarial example
$\mathbf{x}_{\text{adv}}=\mathbf{x}+\boldsymbol{\delta}$
such that $f_v(\mathbf{x}_{\text{adv}})=\text{bonafide}$, while preserving the perceptual and musical content of the original singing audio.
In practice, this corresponds to an attacker attempting to evade automatic SVDD screening on a platform or forensic pipeline.
Since a real-world attacker usually does not know which detector will be used, the objective is to maximize the expected evasion rate over a distribution of possible victim detectors $\mathcal{V}$:
\begin{equation}
    \max_{\boldsymbol{\delta}}
    \mathbb{E}_{f_v \sim \mathcal{V}}
    \left[
    \mathbb{I}(f_v(\mathbf{x}+\boldsymbol{\delta})=\text{bonafide})
    \right].
\end{equation}
A successful attack should therefore transfer across different SSL backbones, layer aggregation strategies, fine-tuned detection heads, and training configurations.

\mypara{Adversary's Knowledge, Capability, and Constraints}
The adversary aims to modify a deepfake singing audio before release or upload such that it is classified as bonafide by an unknown SVDD system while preserving the listening experience and musical semantics of the original clip.
We consider a transfer-based black-box setting: the adversary does not know the victim detector's architecture, parameters, training data, decision threshold, logits, gradients, intermediate representations, or prediction labels, and does not assume any query access to the victim during attack generation.
Instead, the adversary can only access public resources realistically available to practitioners, including open-source speech SSL foundation models such as Wav2Vec 2.0, HuBERT, WavLM, XLS-R, and UniSpeech, as well as public speech or singing datasets.
Using these resources, the adversary constructs a local surrogate ensemble
$\mathcal{S}=\{f_s\}_{s=1}^{K}$
and performs white-box optimization only on the surrogate models.
We emphasize that the attack surrogates are not replicas of the victim detectors.
In our evaluation, adversarial examples are generated using SSL backbones with lightweight MLP heads, whereas the victim systems use different detection heads, training splits, and fine-tuning configurations.
The attacker never observes victim logits, labels, thresholds, gradients, intermediate features, or query responses during optimization, and the generated adversarial audio is transferred to unknown victim detectors without victim-specific adaptation.

The adversary is allowed to modify the waveform under strict perceptual constraints.
Unlike ordinary speech, singing audio contains melody-driven prosody, pitch trajectories, long sustained vowels, rhythm, vibrato, timbre variation, and possible accompaniment.
A perturbation that noticeably changes lyrics, melody, rhythm, singer identity, or audio quality would fail to serve the attacker's purpose and may be detected by listeners or quality-control filters.
Therefore, the perturbation is constrained in both the waveform and frequency domains:
\begin{equation}
    \|\boldsymbol{\delta}\|_\infty \leq \epsilon
    \quad \text{and} \quad
    \mathcal{P}(\boldsymbol{\delta}, \omega) < M(\mathbf{x}, \omega), \forall \omega,
\end{equation}
where $\mathcal{P}(\cdot,\omega)$ denotes the spectral density at frequency $\omega$, and $M(\mathbf{x},\omega)$ denotes the masking threshold induced by the original singing audio.

\section{Methodology}
\subsection{Key Observation}

\label{sec:key_observation}

We begin by observing a structural property of modern SSL-based SVDD systems.
Before task-specific fine-tuning, speech SSL foundation models are trained to capture general acoustic and semantic regularities from large-scale unlabeled audio, rather than spoof-specific artifacts.
As a result, their representations are highly insensitive to singing deepfake artifacts.
Empirically, when using only the last-layer features of pre-trained SSL models for bonafide/spoof discrimination, the resulting detectors perform close to random guessing, with EERs above $50\%$ across different SSL backbones (see \Cref{tab:last_layer_eer}).
This suggests that the pre-trained SSL space primarily preserves semantic and acoustic content, making it a suitable source for constructing a natural anchor.
In contrast, after fine-tuning on singing deepfake datasets, the detector representation becomes explicitly shaped by spoof-discriminative supervision.
This spoof sensitivity is particularly evident in the intermediate layers of the fine-tuned SSL backbone.
As shown in \Cref{tab:ssl_eer_performance}, SLS detectors using mid-layer representations achieve low EERs across different SSL backbones, ranging from $2.62\%$ to $9.30\%$ on CtrSVDD.
Together with our layer-wise analysis, where moving toward very early or very late layers slightly degrades detection performance, this suggests that intermediate fine-tuned representations capture the most discriminative singing-specific spoof artifacts.
Therefore, SSL-based SVDD systems naturally induce two complementary representation spaces:
the \emph{pre-trained SSL space}, which mainly preserves content-related acoustic semantics, and the \emph{fine-tuned detector space}, especially its intermediate layers, which encode spoof-sensitive artifact evidence.
This separation motivates our instance-wise anchor design.
Therefore, SSL-based SVDD systems naturally induce two complementary representation spaces:
the \emph{pre-trained SSL space}, which captures content-preserving natural semantics, and the \emph{fine-tuned detector space}, which encodes spoof-sensitive artifact evidence.
This separation motivates our instance-wise anchor design.

\begin{figure}[t]
    \centering
    \includegraphics[width=0.48\textwidth]{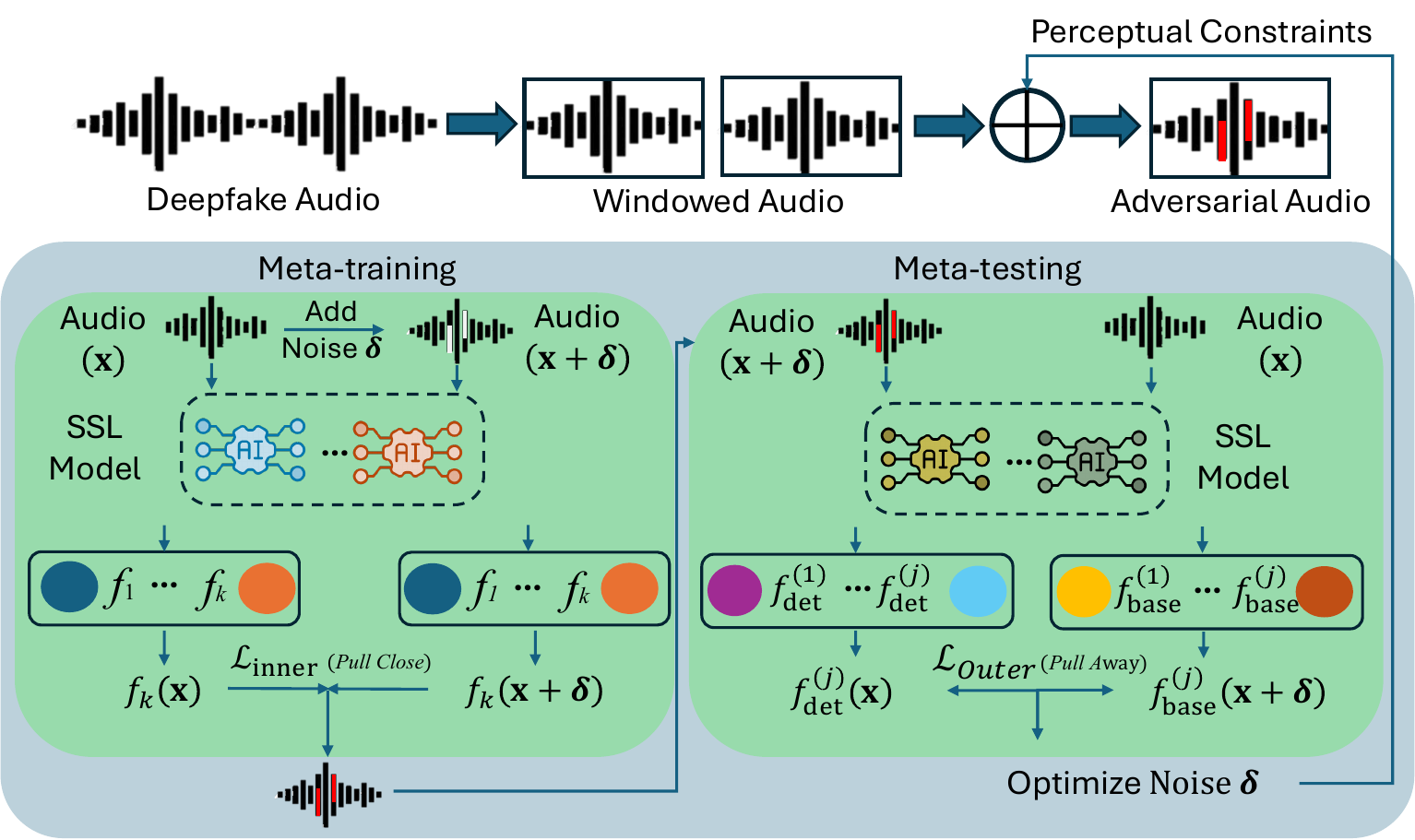}
    \caption{Overview of the proposed Meta-Adversarial Semantic Regression framework. 
    }
    \label{fig:overview}
\end{figure}

\subsection{Approach Overview}
We approach the adversarial generation problem from the perspective of hypothesis testing, which guides our design of the optimization. As shown in ~\Cref{fig:overview}, \method employs a bilevel optimization strategy: the \textbf{Inner Loop} (left) disrupts robust feature correlations in foundation models using an cosine loss, while the \textbf{Outer Loop} (right) employs a ``Push-Pull'' regression to align the adversarial example with the natural  manifold and repel it from the artifact region.

\subsection{Hypothesis-Driven Attack}
\label{sec:method}

\mypara{Attack Inspired by Neyman-Pearson Lemma} 
Let $\mathbf{x} \in \mathbb{R}^T$ be the input audio. 
A generalized deepfake detector can be viewed as comparing evidence between two hypotheses, $H_0$ (bonafide) and $H_1$ (spoof). 
For two fully specified simple hypotheses, the Neyman-Pearson lemma states that the likelihood-ratio test is the most powerful test at a fixed Type I error rate(false alarm).
This motivates us to design an evidence-based attack objective that suppresses spoof-related evidence while preserving bonafide-related semantic evidence.

However, practical SSL-based SVDD detectors do not provide explicit class-conditional densities over raw audio or over a shared representation space.
Moreover, our method uses instance-wise anchors extracted from complementary encoders: a frozen SSL foundation model and fine-tuned surrogate detectors.
Therefore, we do not claim to optimize the exact Neyman-Pearson likelihood ratio.
Instead, we use it as a statistical motivation and construct a local representation-level surrogate objective:
\begin{equation}
    \min_{\boldsymbol{\delta}} \mathcal{L}_{\text{evid}}(\mathbf{x}_{\text{adv}}) 
    =
    \log q_1(\mathbf{z}_{\text{det}}(\mathbf{x}_{\text{adv}})\mid \mathbf{x})
    -
    \log q_0(\mathbf{z}_{\text{base}}(\mathbf{x}_{\text{adv}})\mid \mathbf{x}),
    \label{eq:llr}
\end{equation}
where $q_0$ and $q_1$ denote local evidence models centered at the instance-wise natural and artifact anchors, respectively.
By minimizing this surrogate evidence ratio, the attack encourages $\mathbf{x}_{\text{adv}}$ to move away from detector-specific artifact evidence while remaining aligned with the natural semantic representation.

\mypara{Modeling $H_0$ (Natural Evidence)} 
Following the observation in \Cref{sec:key_observation}, we use the pre-trained SSL space to construct a local source of bonafide-related semantic evidence.
For a deepfake input $\mathbf{x}$, the frozen foundation model extracts a representation that is dominated by acoustic and semantic content and is relatively insensitive to spoof-specific artifacts.
Therefore, we define the instance-wise natural anchor as
$\mathbf{z}_0=\mathbf{z}_{\text{base}}(\mathbf{x})$.
Rather than modeling a global waveform-level density $P(\mathbf{x}|H_0)$, we define a local representation-level evidence model centered at this anchor:
\begin{equation}
   q_0(\mathbf{z}_{\text{base}}(\mathbf{x}_{\text{adv}})\mid \mathbf{x})
   \propto 
   \exp\left(
   \kappa_0 \cdot 
   \mathbf{z}_{\text{base}}(\mathbf{x}_{\text{adv}})^\top \mathbf{z}_0
   \right).
\end{equation}
Maximizing this local evidence encourages the adversarial example to preserve the natural semantic representation, yielding the attractive objective $\mathcal{L}_{\text{pull}}$.

\mypara{Modeling $H_1$ (Artifact Evidence)}
Conversely, fine-tuned detectors are trained to capture synthesis traces that separate bonafide singing from synthetic singing.
The representation of the original deepfake $\mathbf{x}$ in the fine-tuned detector space therefore provides an instance-wise proxy for spoof-related evidence.
We define the artifact anchor as
$\mathbf{z}_1=\mathbf{z}_{\text{det}}(\mathbf{x})$.
Similarly, instead of assuming an explicit global density $P(\mathbf{x}|H_1)$, we define a local artifact-evidence model in the detector representation space:
\begin{equation}
   q_1(\mathbf{z}_{\text{det}}(\mathbf{x}_{\text{adv}})\mid \mathbf{x})
   \propto 
   \exp\left(
   \kappa_1 \cdot 
   \mathbf{z}_{\text{det}}(\mathbf{x}_{\text{adv}})^\top \mathbf{z}_1
   \right).
\end{equation}
Minimizing this local evidence pushes the adversarial example away from the original spoof-sensitive representation, yielding the repulsive objective $\mathcal{L}_{\text{push}}$.
% \section{Meta-Adversarial Optimization}

\begin{figure}[t]
    \centering
    \includegraphics[width=0.95\linewidth]{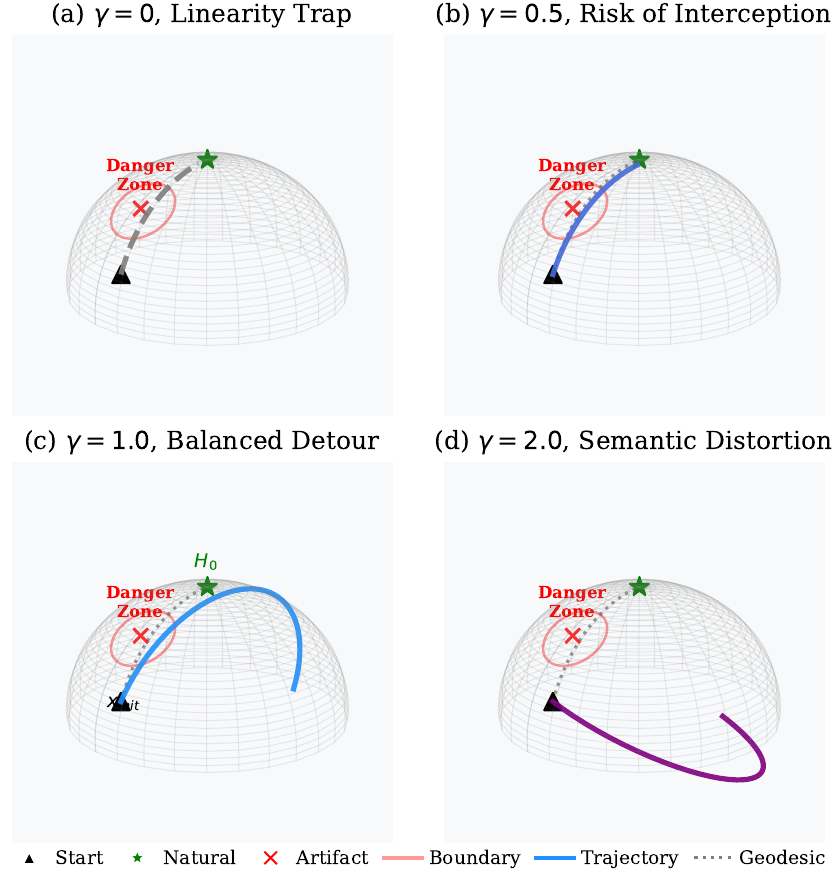}
    \caption{{
    Optimization Dynamics on $\mathbb{S}^2$ under varying equilibrium factor $\gamma$,} with higher $\gamma$ representing higher repulsion.}
    \label{fig:simulation}
\end{figure}

\subsection{Detour in Optimization}
\mypara{The Linearity Trap as Unstable Direct Steering}
\label{sec:geometric_motivation}
A naive joint optimization of 
$\mathcal{L}_{\text{total}}=\mathcal{L}_{\text{pull}}+\gamma\mathcal{L}_{\text{push}}$
updates the perturbation using a single composite gradient.
Although the pull and push terms are defined in complementary representation spaces,
their gradients are both back-propagated to the same waveform perturbation
$\boldsymbol{\delta}$.
In practice, direct joint steering can be sensitive to local interference between
semantic preservation and artifact suppression, yielding a less stable
representation trajectory and weaker black-box transfer.

To make this behavior measurable, we define the waveform-space descent
directions
\begin{equation}
    \mathbf{d}_{\text{pull}}
    =
    -\nabla_{\boldsymbol{\delta}}\mathcal{L}_{\text{pull}},
    \qquad
    \mathbf{d}_{\text{push}}
    =
    -\nabla_{\boldsymbol{\delta}}\mathcal{L}_{\text{push}}.
\end{equation}
We measure their directional interaction by
\begin{equation}
    A_t^{\text{pp}}
    =
    \cos(\mathbf{d}_{\text{pull}},\mathbf{d}_{\text{push}})
    =
    \frac{
    \mathbf{d}_{\text{pull}}^\top \mathbf{d}_{\text{push}}
    }{
    \|\mathbf{d}_{\text{pull}}\|_2
    \|\mathbf{d}_{\text{push}}\|_2
    }.
\end{equation}
A strongly negative value indicates conflict between the two evidence directions,
whereas a value closer to zero indicates weaker directional interference.
We refer to the resulting instability of direct joint steering as the
\emph{Linearity Trap}.

This definition does not require the natural and artifact anchors to lie on the
same representation manifold.
Rather, it characterizes the effective optimization behavior after both losses
are mapped back to waveform space through their corresponding encoders.
The low-dimensional sphere in \Cref{fig:simulation} is therefore only a
didactic illustration of this phenomenon.
In \Cref{tab:trajectory_analysis}, we further quantify this behavior in real SSL systems
using pull-push gradient interaction, inner-joint orthogonality,
representation-space curvature, and geodesic deviation.

To avoid unstable direct steering, the attack should introduce a non-redundant
transverse component while still maintaining a smooth representation trajectory.
This motivates our bi-level look-ahead strategy: the inner step first provides a
transverse semantic update, and the outer step then applies evidence-based
steering from this explored state.

\subsection{Bi-Level Meta-Adversarial Framework}
\label{sec:bilevel}

We propose a bi-level look-ahead optimization strategy to encourage such a
\emph{smooth detour}.
Instead of relying on a single composite gradient at the current point, \method
decomposes the update into two phases.
The inner phase explores a direction that perturbs the base SSL representation,
while the outer phase evaluates the evidence objective at the explored point
and steers the adversarial example toward lower spoof-related evidence.

\begin{itemize}[leftmargin=*]
    \item \textit{Inner Loop (Tangential Repulsion).} 
To escape the predictable gradient path, we force the adversarial example to deviate from its initial semantic representation while staying on the hypersphere. 
With the natural anchor $\mathbf{z}_0$, we formulate it as {minimizing} their cosine similarity:
\begin{equation}
    \min_{\mathbf{\delta}} \mathcal{L}_{\text{tan}}(\mathbf{\delta}) = \mathbf{z}(\mathbf{x} + \mathbf{\delta})^\top \mathbf{z}_0.
\end{equation}
Let $\mathbf{z} = \mathbf{z}(\mathbf{x} + \mathbf{\delta})$ denote the current state. 
The gradient descent update implicitly projects the optimization force onto the tangent space. Specifically, applying the chain rule yields:
\begin{equation}
    \nabla_{\mathbf{\delta}} \mathcal{L}_{\text{tan}} \propto \mathbf{J}^\top (\mathbf{I} - \mathbf{z}\mathbf{z}^\top) \mathbf{z}_0,
\end{equation}
where $\mathbf{J}$ is the Jacobian of the encoder. The projection term $(\mathbf{I} - \mathbf{z}\mathbf{z}^\top)$ ensures that the repulsive force from $\mathbf{z}_0$ is orthogonal to the radial direction $\mathbf{z}$ and introduces a tangent-space component in the normalized SSL representation.
\item \textit{Outer Loop (Evidence Steering).} 
After the inner look-ahead step, we evaluate the local representation-level evidence surrogate at the explored perturbation $\boldsymbol{\delta}'$:
\begin{equation}
\mathcal{L}_{\text{evid}}(\boldsymbol{\delta}')
=
-\ \underbrace{\kappa_0\,\mathbf{z}_{\text{base}}(\mathbf{x}+\boldsymbol{\delta}')^\top \mathbf{z}_0}_{\text{Semantic evidence preservation}}
+
\,\underbrace{\kappa_1\,\mathbf{z}_{\text{det}}(\mathbf{x}+\boldsymbol{\delta}')^\top \mathbf{z}_1}_{\text{Artifact evidence suppression}}.
\end{equation}

Here, the first term encourages the adversarial example to remain aligned with the instance-wise natural anchor in the frozen SSL space, while the second term penalizes similarity to the original spoof-sensitive representation in the fine-tuned detector space.
This objective should be understood as a local evidence surrogate rather than an exact Neyman-Pearson likelihood ratio over raw audio.
To balance the two evidence terms, we define ${\gamma}=\kappa_1/\kappa_0$.
Since the global scaling factor $\kappa_0$ can be absorbed into the optimizer's step size, the effective objective simplifies to:
\begin{equation}
\mathcal{L}_{\text{evid}}
\equiv\;
-\mathbf{z}_{\text{base}}(\mathbf{x}+\boldsymbol{\delta}')^\top \mathbf{z}_0
+{\gamma}\, \mathbf{z}_{\text{det}}(\mathbf{x}+\boldsymbol{\delta}')^\top \mathbf{z}_1.
\end{equation}
Finally, the perturbation is updated using the gradient computed at the explored state $\boldsymbol{\delta}'$:
\begin{equation}
\boldsymbol{\delta}
\leftarrow
\boldsymbol{\delta}
-
\beta
\nabla_{\boldsymbol{\delta}'}
\mathcal{L}_{\text{evid}}(\mathbf{x}+\boldsymbol{\delta}').
\end{equation}
This look-ahead update does not seek a larger path deviation.
Instead, it evaluates the evidence gradient at a transversely explored state,
producing a \emph{smooth detour} that stabilizes the representation trajectory
for black-box transfer.

\end{itemize}

\mypara{Geometric Realization}
To illustrate our optimization intuition, we visualize simplified dynamics on a
low-dimensional unit sphere $\mathbb{S}^2$ (see \Cref{fig:simulation}).
We define a start point $\mathbf{x}_{\text{init}}$ and two illustrative anchors:
the artifact anchor $\mathbf{z}_1$ and the natural anchor $\mathbf{z}_0$.
Under direct optimization ($\gamma=0$), the trajectory tends to follow a short
and locally direct path from the current point.
In this simplified setting, such a path can pass close to the artifact region,
reflecting the potential instability of direct joint steering under pull-push
interference.
In contrast, under the balanced setting of MARS ($\gamma=1$), the trajectory
first explores a transverse semantic direction and then is steered by the
evidence objective, producing a smoother curvilinear trajectory.
Importantly, the purpose of this mechanism is not to maximize path length or
trajectory deviation.
Rather, it produces a \emph{smooth detour}: a non-redundant transverse update
followed by stable evidence steering, which improves transferability while
maintaining a smoother representation trajectory.
This toy visualization should therefore be interpreted as an illustration of the
proposed mechanism rather than as proof of a shared geodesic structure.
In the experiments, we validate this behavior directly in real high-dimensional
SSL systems using trajectory curvature, geodesic deviation, inner-joint
orthogonality, and pull-push gradient interaction.

\begin{algorithm}[t]
\caption{Meta-Adversarial Semantic Regression (MARS)}
\label{alg:mars}
\begin{algorithmic}[1]

\STATE \textbf{Input:} Deepfake audio $\mathbf{x}$, surrogate ensemble $\mathcal{S}=\{f_s\}$, foundation model $f_{\mathrm{base}}$.
\STATE \textbf{Parameters:} Budget $\epsilon$, spectral threshold $\tau_{\mathrm{db}}$, steps $N$, inner/outer rates $\alpha,\beta$, anchor-balance coefficient $\gamma$.

\STATE \textbf{Pre-processing:}
\STATE \parbox[t]{0.88\linewidth}{$
\mathbf{M}_{f,t}
\gets
\mathbb{I}\!\left(
20\log_{10}(|\mathrm{STFT}(\mathbf{x})_{f,t}|)
>
\tau_{\mathrm{db}}
\right)
$}

\STATE \parbox[t]{0.88\linewidth}{$
\mathbf{z}_0
\gets
\mathbf{z}_{\mathrm{base}}(\mathbf{x}),
\quad
\mathbf{z}_1
\gets
\frac{1}{|\mathcal{S}|}
\sum_{f_s\in\mathcal{S}}
\mathbf{z}_{s}(\mathbf{x})
$}

\STATE $\boldsymbol{\delta}_0
\gets
\mathrm{Uniform}(-\epsilon,\epsilon)\odot \mathbf{M}$

\FOR{$n=0$ to $N-1$}

    \STATE \textit{// Phase I: Transverse Look-ahead}

    \STATE $\mathbf{x}_{\mathrm{adv}}
    \gets
    \mathbf{x}
    +
    \boldsymbol{\delta}_n$

    \STATE $\mathcal{L}_{\mathrm{tan}}
    \gets
    \mathbf{z}_{\mathrm{base}}(\mathbf{x}_{\mathrm{adv}})^\top
    \mathbf{z}_0$

    \STATE $\mathbf{g}_{\mathrm{tan}}
    \gets
    \nabla_{\boldsymbol{\delta}_n}
    \mathcal{L}_{\mathrm{tan}}$

    \STATE \parbox[t]{0.88\linewidth}{$
    \boldsymbol{\delta}'
    \gets
    \mathrm{Clip}
    \left(
    \boldsymbol{\delta}_n
    -
    \alpha
    (\mathbf{g}_{\mathrm{tan}}\odot \mathbf{M}),
    -\epsilon,
    \epsilon
    \right)
    $}

    \STATE \textit{// Phase II: Anchor-Conditioned Evidence Steering}

    \STATE $\mathbf{x}'_{\mathrm{adv}}
    \gets
    \mathbf{x}
    +
    \boldsymbol{\delta}'$

    \STATE \parbox[t]{0.88\linewidth}{$
    \begin{aligned}
    \mathcal{L}_{\mathrm{anchor}}
    \gets{}&
    -
    \mathbf{z}_{\mathrm{base}}(\mathbf{x}'_{\mathrm{adv}})^\top
    \mathbf{z}_0 \\
    &+
    \gamma
    \Bigl(
    \frac{1}{|\mathcal{S}|}
    \sum_{f_s\in\mathcal{S}}
    \mathbf{z}_s(\mathbf{x}'_{\mathrm{adv}})^\top
    \mathbf{z}_1
    \Bigr)
    \end{aligned}
    $}

    \STATE $\mathbf{g}_{\mathrm{anchor}}
    \gets
    \nabla_{\boldsymbol{\delta}'}
    \mathcal{L}_{\mathrm{anchor}}$

    \STATE \textit{// Final masked projection update}

    \STATE \parbox[t]{0.88\linewidth}{$
    \boldsymbol{\delta}_{n+1}
    \gets
    \mathrm{Clip}
    \left(
    (
    \boldsymbol{\delta}_n
    -
    \beta
    \mathbf{g}_{\mathrm{anchor}}
    )
    \odot
    \mathbf{M},
    -\epsilon,
    \epsilon
    \right)
    $}

\ENDFOR

\STATE \textbf{Output:} Adversarial example
$\mathbf{x}_{\mathrm{adv}}
=
\mathbf{x}
+
\boldsymbol{\delta}_N$

\end{algorithmic}
\end{algorithm}

\subsection{Perceptual Constraint}
To ensure the generated perturbations remain imperceptible to the human auditory system and bypass standard voice activity filters~\cite{ball2023voice}, we impose a strict frequency-domain constraint utilizing a \textit{Dynamic Spectral Mask}. 
% Unlike standard $\ell_\infty$ constraints that apply noise uniformly, 
We leverage the auditory masking effect by confining perturbations strictly to frequency bands with high spectral energy. 
Let $\mathbf{M} \in \{0, 1\}^{F \times T}$ be a binary mask derived from the Short-Time Fourier Transform (STFT) of the input $\mathbf{x}$, where a time-frequency bin $(f, t)$ is active only if its magnitude exceeds a dynamic threshold $\tau_{\text{db}}$ (set to $-70$ dB in our implementation): $\mathbf{M}_{f,t} = \mathbb{I}(20 \log_{10}(|\text{STFT}(\mathbf{x})_{f,t}|) > \tau_{\text{db}})$. As enforced in our optimization algorithm, this mask is applied at three critical stages: during the \textit{initialization} of the random perturbation ($\boldsymbol{\delta} \leftarrow \boldsymbol{\delta} \odot \mathbf{M}$), to the \textit{gradient updates} ($\nabla_{\boldsymbol{\delta}} \leftarrow \nabla_{\boldsymbol{\delta}} \odot \mathbf{M}$) in both inner and outer loops, and for the \textit{final projection} after every update step.
This guarantees that the optimization direction never explores perceptible low-energy regions, allowing for potent adversarial manipulation in salient frequency bands while keeping the fidelity.

\subsection{Final Approach}
\label{sec:algorithm_summary}

Algorithm~\ref{alg:mars} summarizes the overall procedure of \method.
Given a deepfake audio sample $\mathbf{x}$ and a surrogate ensemble $\mathcal{S}$ (Line~1), \method first specifies the perturbation budget, spectral threshold, optimization steps, learning rates, and anchor-balance coefficient (Line~2). In the pre-processing stage (Line~3), it computes a dynamic spectral mask $\mathbf{M}$ from the STFT magnitude of the input audio (Line~4), so that perturbations are restricted to high-energy time-frequency regions. It then extracts two instance-wise semantic anchors (Line~5): the natural anchor $\mathbf{z}_0$ from the foundation model $f_{\mathrm{base}}$, and the artifact anchor $\mathbf{z}_1$ from the surrogate ensemble $\mathcal{S}$. The perturbation $\boldsymbol{\delta}_0$ is initialized uniformly within the $\ell_\infty$ budget and masked by $\mathbf{M}$ (Line~6). During optimization, each iteration (Line~7) consists of two stages. In the transverse look-ahead stage (Line~8), \method constructs the current adversarial sample (Line~9), computes the base-model similarity loss with respect to $\mathbf{z}_0$ (Line~10), and obtains its gradient $\mathbf{g}_{\mathrm{tan}}$ (Line~11). A masked and clipped update then produces the intermediate perturbation $\boldsymbol{\delta}'$ (Line~12), encouraging exploration away from the direct semantic direction. In the anchor-conditioned evidence steering stage (Line~13), \method evaluates the look-ahead adversarial sample $\mathbf{x}'_{\mathrm{adv}}$ (Line~14) and optimizes the anchor objective (Line~15), where the base-model term guides the sample toward the natural semantic anchor while the surrogate-ensemble term, weighted by $\gamma$, suppresses artifact-related evidence. The corresponding gradient $\mathbf{g}_{\mathrm{anchor}}$ is then computed (Line~16). Finally, \method performs the masked projection update (Line~17) by updating the original perturbation with $\mathbf{g}_{\mathrm{anchor}}$, applying the spectral mask, and clipping it to the $\ell_\infty$ budget (Line~18). After $N$ iterations, the final adversarial example is returned as $\mathbf{x}_{\mathrm{adv}}=\mathbf{x}+\boldsymbol{\delta}_N$ (Line~20).

\section{Attack Evaluation}
\subsection{Evaluation Setup}
\mypara{Dataset}
To evaluate the effectiveness of our adversarial attack on singing voice deepfake detection, we use the CtrSVDD dataset~\cite{DBLP:conf/slt/ZhangZSYTD24} as the testing dataset. This dataset includes 7 SVS and 7 SVC methods to generate 188,486 (260.34 hours) deepfake song clips against 32,312 (47.64 hours) bonafide song clips for 164 singers in different languages~\cite{zhang2024xwsb,liu2025nes2net}.

\mypara{Target Model}
Since current SVDD methods typically use a ``pre-trained SSL model plus fine-tuned head'' paradiagm, we select 8 representative SSL foundation models and 3 detecting heads.
Specifically, for SSL foundation models, we have WavLM (Base, Large)~\cite{chen2022wavlm}, HuBERT (Base, Large)~\cite{hsu2021hubert}, Wav2Vec 2.0 (Base, XLSR)~\cite{baevski2020wav2vec,babu2021xls}, and Unispeech-SAT (Base, Large)~\cite{chen2022unispeech}. 
Regarding the detecting head, we have AASIST2~\cite{jung2022aasist}, SLS~\cite{zhang2024audio}, and MultiConv~\cite{tran2025multi}. 
We combine the SSL models and detection heads as the black-box target models. 
All of them are trained on the \textit{training set} of the CtrSVDD dataset following their official implementations.

\mypara{Baselines}
For fair comparison, we evaluate \method against a comprehensive set of competitive baseline attacks under the standard ensemble-based black-box setting~\cite{liu2016delving}.
We organize them into three groups according to their attack objectives and transfer mechanisms.

\textit{Surrogate-boundary optimization baselines.}
The first group contains attacks that directly optimize the surrogate detector's classification objective.
We include Ensemble Projected Gradient Descent (PGD)~\cite{uddin2025adversarial} and Ensemble C\&W~\cite{uddin2025adversarial}.
PGD iteratively maximizes the surrogate cross-entropy loss under an $\ell_\infty$ constraint, while C\&W optimizes a margin/confidence-based objective.
Both methods mainly seek perturbations that cross the surrogate decision boundary, and therefore instantiate the boundary-centric transfer setting discussed in \Cref{sec:geometric_motivation}.

\textit{Transfer-enhanced gradient baselines.}
The second group includes representative transferable attacks that improve black-box transfer by modifying the gradient update, input transformation, or surrogate model.
We evaluate MI-FGSM~\cite{dong2018boosting}, which stabilizes iterative updates with momentum; DI-FGSM~\cite{xie2019improving}, which applies random input transformations to reduce overfitting to a single input view; TI-FGSM~\cite{dong2019evading}, which smooths gradients to encourage translation-invariant perturbations; SI-NI-FGSM~\cite{lin2019nesterov}, which combines scale invariance with Nesterov accelerated gradients; VMI-FGSM~\cite{wang2021variance}, which reduces local gradient variance for more stable transferable updates; and AWT~\cite{chen2025awt}, which adaptively tunes surrogate weights during attack generation.
Although these methods substantially enhance transferability, they still primarily refine the surrogate optimization process and tend to follow the dominant surrogate gradient direction, corresponding to the gradient-direction bias discussed in \Cref{sec:geometric_motivation}.

\textit{Direct Push-Pull objective baseline.}
To empirically validate the geometric analysis in \Cref{sec:geometric_motivation}, we further include a Joint Optimization baseline.
This baseline directly optimizes our combined Push-Pull objective,
$\mathcal{L}_{\text{pull}}+\mathcal{L}_{\text{push}}$,
using gradient descent without the proposed bi-level detour mechanism.
Comparing \method with this baseline isolates the benefit of Tangential Repulsion and verifies whether escaping the direct geodesic path is necessary for transferability.

\begin{table*}[t]
\centering
\caption{Performance Comparison across Detectors and SSL Models in Black-box Settings (ASR, \%)}
\label{tab:blackbox_comparison_full}
\renewcommand{\arraystretch}{0.9}
\setlength{\tabcolsep}{3pt}
\scriptsize
\vspace{-2mm}
\resizebox{0.9\textwidth}{!}{%
\begin{tabular}{@{} ll cccccccc @{}}
\toprule
\multirow{2}{*}{\textbf{Detector}} & \multirow{2}{*}{\textbf{Method}} & 
\multicolumn{8}{c}{\textbf{SSL Models}} \\
\cmidrule(l){3-10}
 &  & 
\textbf{Wav2vec-B} & \textbf{Wav2vec-X} & \textbf{WavLM-B} & \textbf{WavLM-L} &
\textbf{Hubert-S} & \textbf{Hubert-L} & \textbf{Unispeech-B} & \textbf{Unispeech-L} \\
\midrule

\multirow{10}{*}{\textbf{AASIST2}} 
 & PGD     
 & 47.18 & 19.76 & 70.97 & 66.53 & 53.62 & 29.43 & 55.25 & 45.16 \\
 & MI-FGSM
 & 75.00 & 45.48 & 80.72 & 29.22 & 61.75 & 27.11 & 77.71 & 54.82 \\
 & DI-FGSM 
 & 81.29 & 83.93 & 90.96 & \underline{93.43} & 85.25 & \underline{76.14} & 89.54 & 80.83 \\
 & TI-FGSM
 & 75.23 & 67.65 & 83.59 & 74.05 & 88.29 & 75.69 & 85.84 & 81.57 \\
 & AWT
 & 78.50 & 82.63 & 85.38 & 75.94 & 68.49 & 65.83 & 76.35 & 73.40 \\
 & SI-NI-FGSM
 & 73.85 & 75.94 & 80.77 & 80.63 & 76.29 & 73.85 & 80.70 & 72.88 \\
 & VMI-FGSM
 & \underline{84.15} & \underline{88.73} & 85.58 & 93.27 & 87.50 & 75.19 & 85.58 & \underline{82.88} \\
 & C\&W 
 & 83.48 & 87.92 & \underline{93.33} & 80.01 & \underline{90.42} & 74.38 & \underline{90.28} & 74.20 \\
\cmidrule{2-10}
 & Joint
 & 74.59 & 49.34 & 82.28 & 80.60 & 74.92 & 51.16 & 77.39 & 79.75 \\ 
 & \method  
 & \textbf{90.35} & \textbf{92.77} & \textbf{95.18} & \textbf{98.80} & \textbf{91.07} & \textbf{81.70} & \textbf{91.19} & \textbf{92.17} \\ 
\midrule

\multirow{10}{*}{\textbf{SLS}} 
 & PGD
 & 29.83 & 10.08 & 43.55 & 56.19 & 59.67 & 43.56 & 59.67 & 69.37 \\
 & MI-FGSM 
 & 40.96 & 13.86 & 42.47 & 17.17 & 71.69 & 12.95 & 51.51 & 55.12 \\
 & DI-FGSM
 & 70.24 & 54.68 & 74.31 & 71.64 & 74.17 & 58.02 & 85.17 & 73.70 \\
 & TI-FGSM
 & 62.24 & 46.83 & 75.16 & 71.08 & 75.47 & 52.05 & 80.21 & 72.46 \\
 & AWT
 & 66.27 & \underline{62.63} & 75.38 & 65.94 & 68.49 & \underline{65.83} & 76.35 & 73.40 \\
 & SI-NI-FGSM
 & 62.63 & 52.14 & 67.76 & 64.91 & 70.37 & 49.48 & 75.32 & 71.57 \\
 & VMI-FGSM
 & 68.52 & 53.46 & 73.79 & 66.35 & 76.63 & 63.44 & 88.57 & 79.95 \\
 & C\&W 
 & \underline{74.58} & 56.67 & \underline{87.08} & 51.88 & \underline{92.71} & 50.63 & \underline{89.79} & \underline{82.92} \\
\cmidrule{2-10}
 & Joint
 & 70.19 & 55.37 & 77.56 & \underline{69.63} & 68.12 & 48.67 & 71.43 & 68.17 \\
 & \method  
 & \textbf{77.23} & \textbf{85.90} & \textbf{92.17} & \textbf{95.42} & \textbf{95.18} & \textbf{80.48} & \textbf{93.63} & \textbf{97.59} \\
\midrule

\multirow{10}{*}{\textbf{MultiConv}} 
 & PGD    
 & 62.58 & 14.18 & 74.60 & 49.19 & 68.15 & 27.17 & 66.53 & 50.83 \\
 & MI-FGSM 
 & 75.04 & 17.83 & 74.70 & 15.12 & 73.80 & 12.71 & 61.45 & 20.48 \\
 & DI-FGSM
 & 85.51 & 43.37 & 70.79 & 64.26 & 71.88 & \underline{52.67} & 66.50 & 57.10 \\
 & TI-FGSM
 & 75.67 & 48.50 & 72.48 & 68.82 & 68.56 & 53.66 & 72.48 & 66.81 \\
 & AWT
 & 68.37 & 53.92 & 74.18 & 63.16 & 78.10 & 52.11 & 76.34 & 68.95 \\
 & SI-NI-FGSM
 & 66.36 & 51.75 & 63.08 & 56.89 & 76.57 & 41.12 & 73.29 & 69.79 \\
 & VMI-FGSM
 & \underline{89.81} & \underline{55.38} & 78.46 & \underline{76.03} & 79.85 & 48.62 & 80.77 & 68.25 \\
 & C\&W 
 & 85.42 & 47.08 & \underline{91.88} & 67.13 & \underline{92.96} & 27.50 & \underline{91.46} & 55.83 \\
\cmidrule{2-10}
 & Joint
 & 42.74 & 45.43 & 75.74 & 56.92 & 91.67 & 30.53 & 76.37 & \underline{71.35} \\ 
 & \method  
 & \textbf{90.36} & \textbf{80.83} & \textbf{93.37} & \textbf{92.17} & \textbf{95.48} & \textbf{54.79} & \textbf{94.75} & \textbf{92.05} \\
\bottomrule

\end{tabular}
}
\end{table*}

\mypara{Metrics}
We adopt the Attack Success Rate (ASR) to evaluate the evasion capability of our attack.
Let $\mathbb{X}_{\text{hit}}$ denote the subset of deepfake samples correctly identified as spoof by the victim detector $D(\cdot)$. The ASR is defined as the percentage of these samples that are misclassified as bonafide after perturbation:
\begin{equation}
     \text{ASR} = \frac{1}{|\mathbb{X}_{\text{hit}}|} \sum_{\mathbf{x} \in \mathbb{X}_{\text{hit}}} \mathbb{I}(D(\mathbf{x} + \boldsymbol{\delta}) = \text{bonafide}),
\end{equation}
where $\mathbb{I}(\cdot)$ is the indicator function. 
A higher ASR indicates stronger attack potency.

\mypara{Implementation} 
We construct generalized surrogate networks comprising an SSL backbone followed by a simple MLP classification head.
We strictly enforce the black-box constraint by training these surrogates on the validation set of the CtrSVDD dataset, while the victim models are trained on the CtrSVDD's training set.
For the specific instantiation of our bi-level framework, we utilize \textbf{Wav2vec-Base} as the inner surrogate model and a fine-tuned \textbf{WavLM-Large} as the outer surrogate model. Guided by recent analyses on the layer-wise attention of deepfake detectors~\cite{el2025comprehensive}, we strategically select feature layers to optimize the trajectory: for the outer steering phase, we align the intermediate features (layers 10-14) of the fine-tuned WavLM-Large towards the semantic representation of the last layer of its frozen foundation model. 
Besides, for the inner exploration phase, we maximize the feature deviation in layers 4-6 of the Wav2vec-Base, pushing them away from their original states. 
The entire optimization process is driven by PGD with a perturbation budget $\epsilon = 1 \times 10^{-2}$, step size $\alpha = 4 \times 10^{-3}$, and total iterations $T = 30$.

As for the baseline implements, PGD follows ours setting. 
For audio-domain baselines that have been previously evaluated on audio spoofing or audio deepfake detection, such as C\&W and MI-FGSM, we directly follow the corresponding audio-domain implementations and parameterization protocols. 
To be specific, for the C\&W baseline, we adopt the optimization-based formulation with a regularization weight $c = 100$ and a confidence margin $\kappa = 0$.
For the MI-FGSM baseline, we incorporate a momentum term with a decay factor $µ = 1.0$ to stabilize the gradient update directions. 
For transferable attacks originally developed in the image domain, including DI-FGSM, TI-FGSM, SI-NI-FGSM, VMI-FGSM, and AWT, we adapt their official implementations to waveform inputs under the same perturbation budget, surrogate access, and perceptual constraints.
Bedsides, all baseline hyperparameters are selected using only the local surrogate models; no victim-side labels, logits, thresholds, or gradients are used for tuning.
\Cref{app:blackbox_baseline_protocol} provides the full black-box separation, baseline fairness, and hyperparameter tuning protocol.

\subsection{Attack Performance}

\mypara{In-distribution Performance}
\Cref{tab:blackbox_comparison_full} reports the black-box transfer performance across three detector architectures and eight SSL backbones.
Overall, \method achieves the strongest performance among all methods: it obtains the highest ASR in all $24$ detector-backbone configurations and exceeds $90\%$ ASR in $18$ of them.
Averaged over all settings, \method reaches an ASR of $89.36\%$, substantially outperforming the strongest transfer-enhanced baselines, including VMI-FGSM ($76.28\%$) and C\&W ($75.81\%$).
The improvement is consistent across detector families: \method achieves average ASRs of $91.65\%$, $89.70\%$, and $86.72\%$ on AASIST2, SLS, and MultiConv, respectively, improving over the best baseline by $6.29$, $16.42$, and $14.58$ percentage points.

The results also reveal the limitations of existing attacks.
Boundary-centric methods such as PGD and MI-FGSM generalize poorly across unseen detectors and SSL backbones.
For example, MI-FGSM drops to $12.71\%$ ASR on MultiConv with HuBERT-L and $17.17\%$ on SLS with WavLM-L, indicating severe overfitting to surrogate decision boundaries.
Transfer-enhanced image-domain attacks improve over PGD and MI-FGSM, but their performance remains unstable: VMI-FGSM achieves strong results on AASIST2 and some MultiConv settings, yet drops to $48.62\%$ on MultiConv with HuBERT-L; C\&W performs well on several backbones but only reaches $27.50\%$ in the same hard setting.
Moreover, Joint Optimization, which directly optimizes the Push-Pull objective without the proposed detour mechanism, still fails on several configurations, e.g., $30.53\%$ on MultiConv with HuBERT-L and $48.67\%$ on SLS with HuBERT-L.
These results support our geometric analysis: directly crossing surrogate boundaries or following the dominant optimization direction is insufficient for robust black-box transfer.
In contrast, \method generates perturbations that are more invariant to detector heads, SSL backbones, and layer aggregation strategies, demonstrating the effectiveness of the proposed bi-level detour mechanism.

\subsection{Out-of-distribution (OOD) Dataset Performance}
\begin{figure}[t]
    \centering
    \includegraphics[width=1\columnwidth]{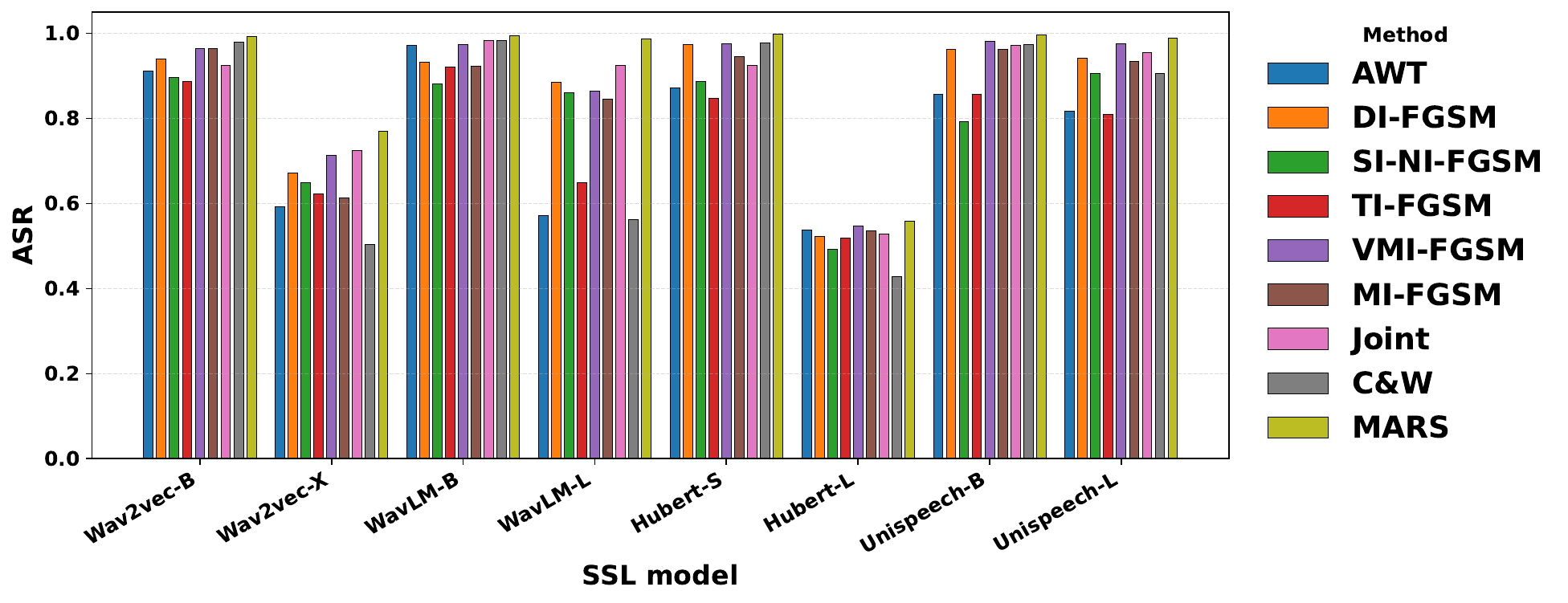}
    % \vspace{-2mm}
    \caption{\small Transferability of adversarial attacks on the
OOD dataset across various SSL backbones on the Multconv. }
    \label{fig:OOD_dataset}
\end{figure}

\Cref{fig:OOD_dataset} evaluates cross-dataset transferability on the FsD~\cite{xie2024fsd} dataset, which introduces distribution shifts in both singing vocal characteristics and forgery methods compared with CtrSVDD. 
Despite these shifts, MARS preserves high ASR across most evaluated detector-backbone configurations. 
For instance, it reaches 99.72\% ASR on AASIST with WavLM-B and 99.87\% ASR on MultiConv with HuBERT-S. 
These results provide empirical evidence that MARS is not limited to the source dataset and can transfer to unseen singing deepfake distributions. 
Nevertheless, we interpret the results as evidence of strong cross-dataset transfer under the evaluated settings, rather than as a claim that the perturbations are universally dataset-invariant.

\subsection{Cross-Domain Generalization }
To assess the adaptability of \method beyond academic benchmarks, we evaluate its robustness against real-world, high-fidelity AI music generated by industrial platforms. 
We utilize the Sonics dataset~\cite{rahman2024sonics}, which comprises full-band songs produced by commercial black-box models such as Udio~\cite{udio2024} and Suno~\cite{suno2024}. 
This setting represents a challenging \textit{cross-domain transfer} scenario, as the generation architectures differ significantly from the SVS/SVC systems in the training set.
Specifically, we employ the Demucs framework~\cite{rouard2022hybrid} to separate the AI-generated vocal tracks and generate adversarial perturbations on Demucs.
Crucially, to validate our geometric hypothesis, we compare \method against the \textit{Joint Optimization} baseline on this setting.

\mypara{Results and Discussion}
As shown in~\Cref{tab:transferability}, \method demonstrates near-perfect generalization on the UniSpeech ($99.13\%$) and WavLM ($97.61\%$) families, maintaining a high average ASR above $80\%$ across all tested architectures. 
In sharp contrast, the Joint Optimization baseline suffers a significant performance drop compared to \method, lagging by $42.27\%$ on average ($86.01\%$ vs. $43.74\%$). 
Specifically, while \method achieves an average ASR of $86.23\%$ on the HuBERT family, the baseline fails to generalize, yielding only $39.99\%$. 
This disparity provides additional empirical support for the direct-steering bias hypothesis in an OOD setting, where optimization along a direct geodesic path in a specific surrogate's feature space results in severe overfitting. By enforcing a curvilinear detour via the Bi-Level mechanism, \method effectively decouples the attack from source-specific artifact patterns, suggesting that tangential exploration can improve cross-distribution robustness when combined with evidence steering.

\begin{table}[h]
\centering
\caption{Transferability of adversarial attacks on the Sonics' AI Instruments dataset across various SSL backbones. We compare our method against the Joint Optimization baseline.}
\label{tab:transferability}
\footnotesize
\resizebox{0.7\linewidth}{!}{%
\renewcommand{\arraystretch}{0.8}
\begin{tabular}{llcc}
\toprule
\multirow{2}{*}{\textbf{Model Family}} & \multirow{2}{*}{\textbf{Variant}} & \multicolumn{2}{c}{\textbf{ASR (\%)}} \\
\cmidrule(l){3-4}
 &  & Joint Opt. & \textbf{Ours} \\ \midrule
\multirow{2}{*}{HuBERT} & Large & 38.65 & \textbf{84.56} \\
 & Small & 41.32 & \textbf{87.90} \\ \midrule
\multirow{2}{*}{WavLM} & Large & 51.56 & \textbf{95.61} \\
 & Base & 48.06 & \textbf{92.09} \\ \midrule
\multirow{2}{*}{UniSpeech} & Large & 53.45 & \textbf{94.13} \\
 & Base & 49.37 & \textbf{93.07} \\ \midrule
\multirow{2}{*}{Wav2Vec 2.0} & XLSR & 65.94 & \textbf{82.63} \\
 & Base & 64.32 & \textbf{78.13} \\ \bottomrule
\end{tabular}%
}
\end{table}

\mypara{Optimization Dynamics Analysis}
To examine whether MARS changes the optimization behavior, we analyze
the attack trajectories on 500 adversarial examples sampled from the test set.
This analysis is conducted offline after attack generation and is not used by
the attacker during optimization.
We compute the trajectory statistics in the real high-dimensional SSL
representation space.
We report trajectory curvature and geodesic deviation as smoothness and
stability metrics, where lower values indicate smoother and more stable
representation steering. We also report
$\cos(\mathbf{d}_{\text{pull}},\mathbf{d}_{\text{push}})$ to measure the
directional interaction between semantic preservation and artifact suppression.

As shown in~\Cref{tab:trajectory_analysis}, \method produces the lowest curvature and near-lowest geodesic deviation among the compared attacks.
Compared with Joint Optimization, MARS reduces trajectory curvature from
$2.43$ to $2.03$ and lowers mean geodesic deviation from $0.173$ to $0.119$.
Although VMI-FGSM obtains a slightly lower geodesic deviation, it has much lower
ASR and higher curvature, suggesting that low deviation alone is insufficient
for transferability. MARS provides the best trade-off between attack success and
trajectory stability.
In addition, MARS yields the weakest pull-push conflict,
with $\cos(\mathbf{d}_{\text{pull}},\mathbf{d}_{\text{push}})=-0.197$, which is
closer to zero than the other baselines. These results support our claim that
MARS improves transferability by inducing a smooth detour: a transverse
look-ahead step followed by stable evidence-based steering, rather than by
simply increasing trajectory length.

\mypara{Mechanism Ablation} To verify that the proposed detour is not merely a stochastic perturbation or a by-product of the Push-Pull objective, we compare MARS with three controlled variants: random detour followed by Push-Pull, inner-only tangential exploration, and outer-only evidence steering. The results show that random detours are insufficient, achieving only 75.18\% average ASR, which is 11.55 percentage points lower than full MARS. This gap is particularly large on difficult transfer targets such as Wav2vec-X and UniSpeech-L, where MARS improves over random detour by 43.50 and 28.59 percentage points, respectively. Inner-only also underperforms MARS by 9.16 percentage points on average, indicating that tangential exploration alone cannot reliably evade detectors without subsequent evidence steering. Outer-only is the strongest partial variant, confirming that the Push-Pull evidence objective is important. Besides, full MARS still improves the average ASR from 85.69\% to 86.73\%, with notable gains on Wav2vec-X (+10.16). These results suggest that the benefit of MARS comes from the combination of a learned tangential look-ahead step and evidence-based steering, rather than from arbitrary path perturbation.

\begin{table}[t]
\centering
\caption{
Optimization dynamics of different attacks.
Curvature and geodesic deviation are lower-is-better smoothness/stability
metrics.
$\cos(\mathbf{d}_{\text{pull}},\mathbf{d}_{\text{push}})$ measures the
directional interaction between semantic preservation and artifact suppression;
values closer to zero indicate weaker pull-push conflict.
MARS achieves the highest ASR and the lowest curvature while maintaining
near-lowest geodesic deviation.
}
\label{tab:trajectory_analysis}
\resizebox{\linewidth}{!}{
\begin{tabular}{lcccc}
\toprule
Method 
& Curvature $\downarrow$
& Geo. Dev. $\downarrow$
& $\cos(\mathbf{d}_{\text{pull}},\mathbf{d}_{\text{push}})$ \\
\midrule
Joint 
& 2.43 
& 0.173 
& -0.228 \\
MARS 
& \textbf{2.03} 
& \textbf{0.119}
& \textbf{-0.197} \\
C\&W 
& 2.47 
& 0.152 
& -0.373 \\
VMI-FGSM 
& 2.56 
& 0.123
& -0.418 \\
\bottomrule
\end{tabular}
}
\end{table}

\subsection{Ablation Study}
\mypara{Compatibility of the VMF Assumption}
To validate the theoretical grounding of our hypothesis testing framework, we ablate the probability distribution used to model the feature space.
Specifically, we compare our vMF-based objective (inner product) against Gaussian (MSE/$\ell_2$) and Laplacian ($\ell_1$) priors, which assume a flat Euclidean geometry.
Results in \Cref{fig:loss_ablation} reveal that \textit{alignment with the intrinsic manifold geometry is a prerequisite for transferability.}

Since SSL models trained with contrastive objectives naturally map data onto a hyperspherical manifold $\mathbb{S}^{d-1}$, magnitude-sensitive metrics like MSE and $\ell_1$ suffer from a geometric mismatch in this space.
In the Inner Loop, optimizing MSE induces gradients with significant radial components. 
This wastes optimization budget on changing the vector norm, rather than focusing on the angular deviation required to escape the local optimum. 
In contrast, the vMF-derived Cosine loss generates purely tangential gradients, efficiently driving the adversarial sample into the orthogonal null-space.
Similarly, in the {Outer Loop}, Euclidean constraints prove overly rigid.
They enforce a strict point-to-point distance penalty, penalizing samples that are semantically aligned but differ in magnitude.
The vMF distribution provides the necessary flexibility for the adversary to traverse the manifold curvature. 
The superior performance of Cosine loss confirms that attacking SSL-based detectors requires respecting their hyperspherical nature rather than treating them as flat Euclidean spaces.
\begin{figure}[t]
    \centering
    \begin{subfigure}[t]{0.48\linewidth}
        \centering
        \includegraphics[width=\linewidth]{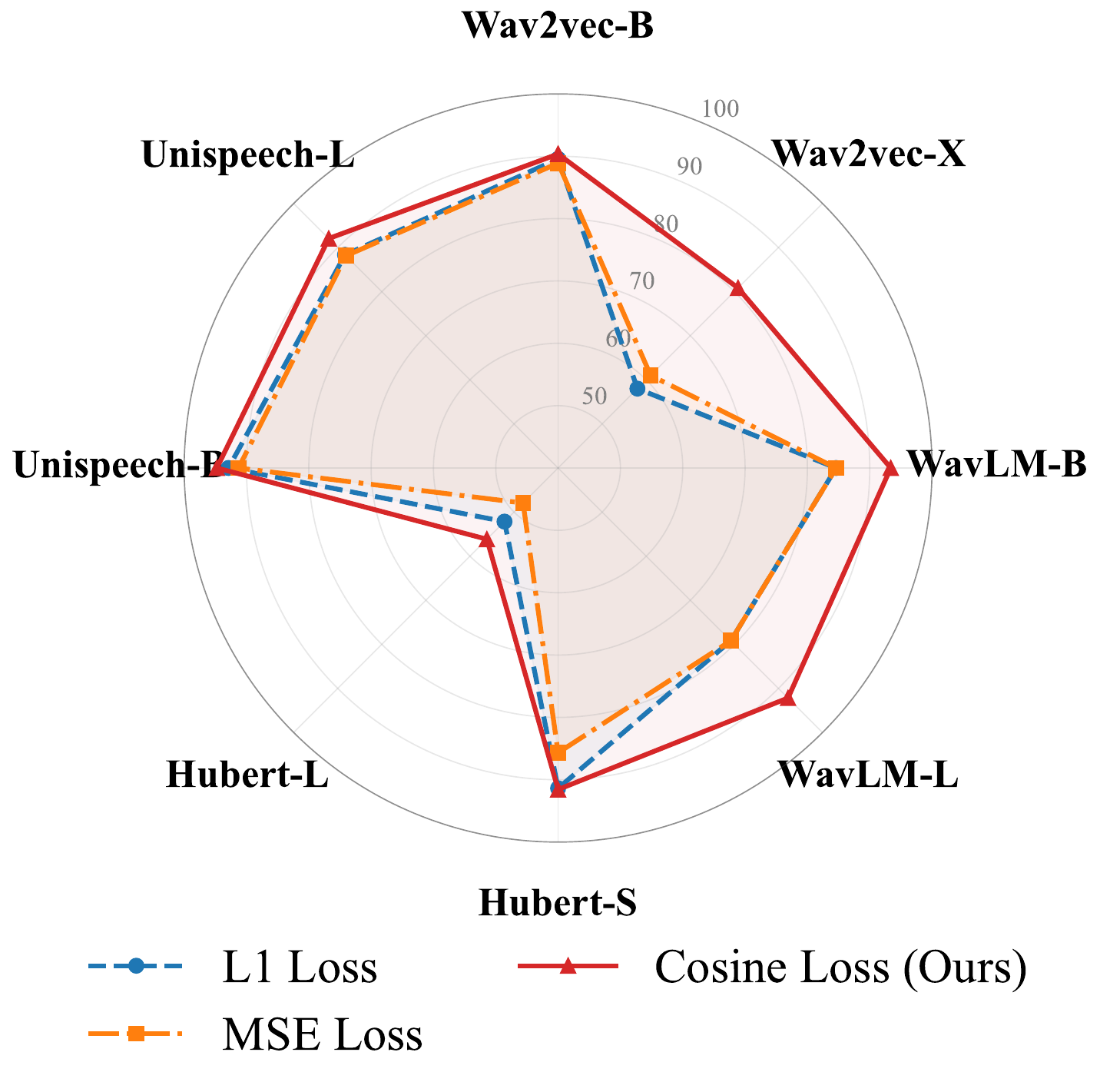}
        \caption{Stage 1: Inner Loop Loss}
        \label{fig:stage1_loss}
    \end{subfigure}
    \hfill
    \begin{subfigure}[t]{0.48\linewidth}
        \centering
        \includegraphics[width=\linewidth]{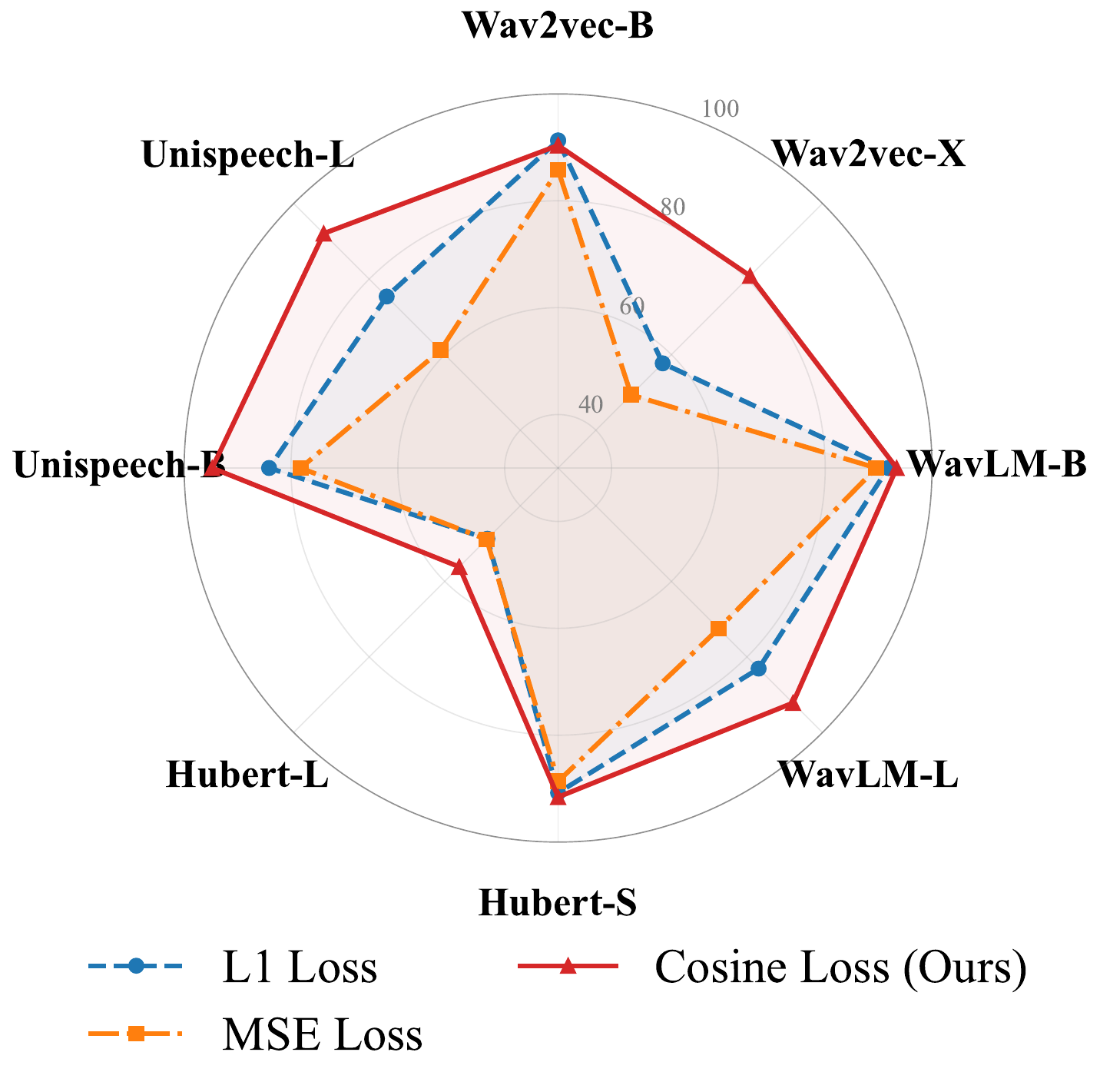}
        \caption{Stage 2: Outer Loop Loss}
        \label{fig:stage2_loss}
    \end{subfigure}
    
    \caption{Ablation study on loss functions for the MultiConv detector. (a) Comparison of loss functions in Stage 1 (Disruption). (b) Comparison of loss functions in Stage 2 (Regression).}
    \label{fig:loss_ablation}
\end{figure}

\begin{figure}[htbp]
    \centering
    \begin{subfigure}[b]{0.48\linewidth}
        \centering
        \includegraphics[width=\linewidth]{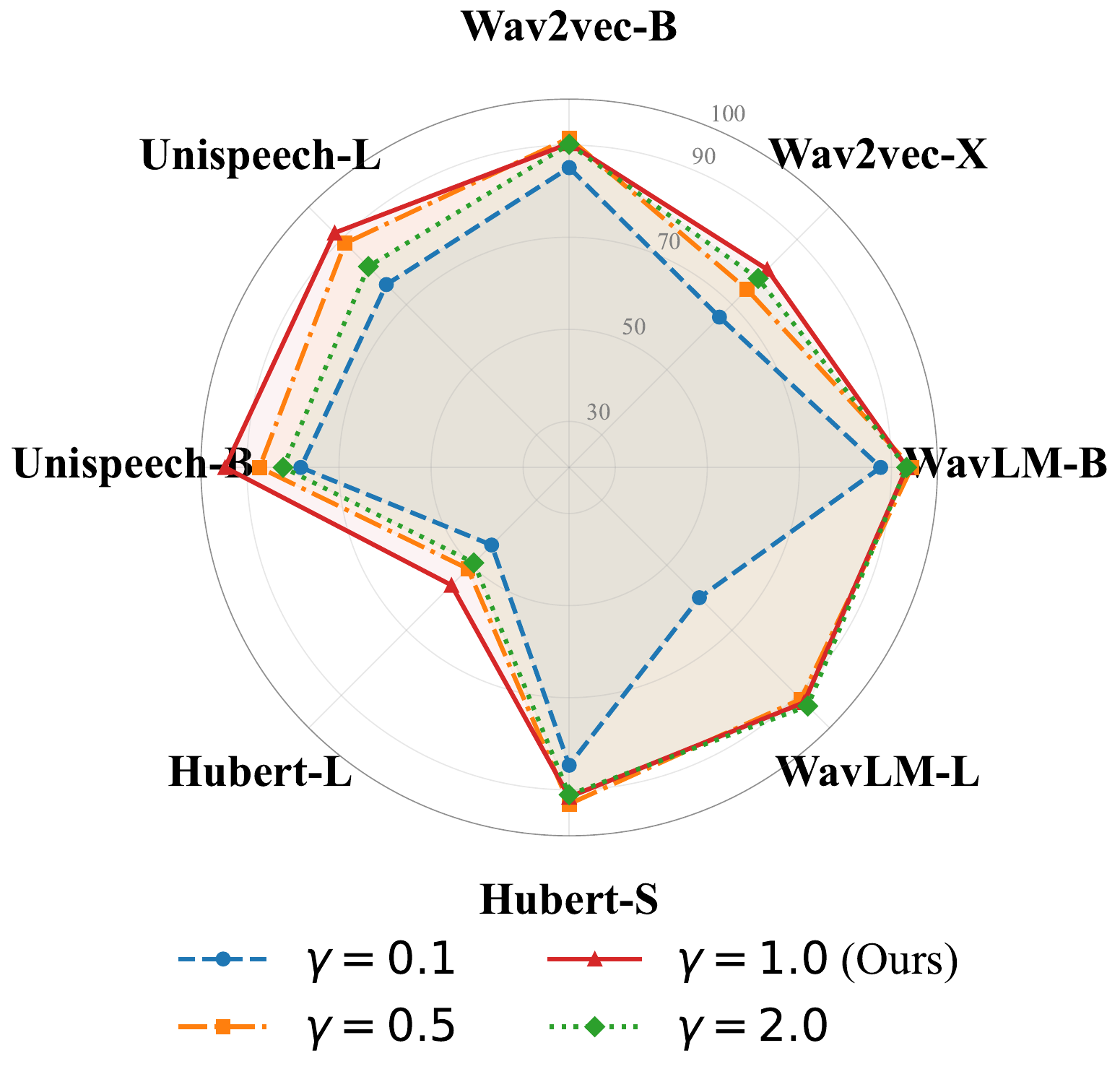}
        \caption{\small Ablation $\gamma$ in Hypothesis Steering.}
        \label{fig:gamma_ablation}
    \end{subfigure}
    \hfill
    \begin{subfigure}[b]{0.48\linewidth}
        \centering
        \includegraphics[width=\linewidth]{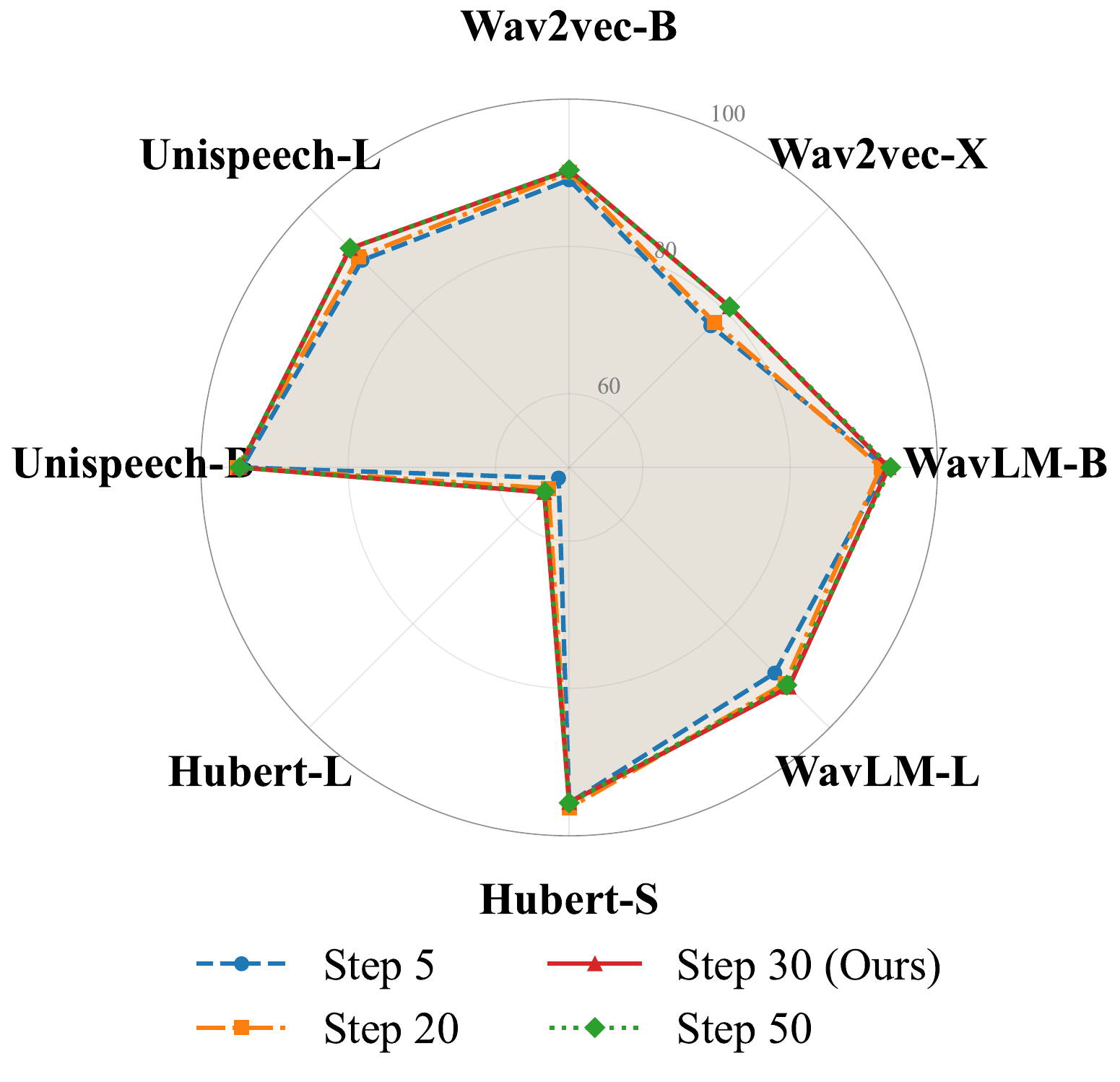}
        \caption{\small Ablation $step$ in Adversary Attack.}
        \label{fig:step_ablation}
    \end{subfigure}
    
    \caption{Ablation study on $\gamma$ and $step$ for the MultiConv detector.}
    \label{fig:loss_ablation}
\end{figure}

\mypara{Equilibrium in Hypothesis Steering}
\label{sec:equilibrium}
We also analyze ${\gamma}$, which governs the ratio of the concentration parameters of the two hypothesis distributions, as a proxy for the equilibrium between semantic preservation (Pull to $H_0$) and artifact suppression (Push from $H_1$).
Specifically, we evaluate the attack on the MultiConv by varying ${\gamma}$, as shown in ~\Cref{fig:gamma_ablation}.
A remarkable observation is that the ASR peaks consistently around ${\gamma} \approx 1.0$. 
This empirical optimality suggests that effective evasion requires a {symmetric contribution} from both objectives: the adversarial example must simultaneously act as a ``denoised'' semantic sample and an ``anti-artifact'' sample.

When ${\gamma} < 1.0$ (see \Cref{fig:simulation}(b)), the optimization is dominated by the semantic pull from $\mathbf{z}_0$. 
As analyzed in ~\Cref{sec:bilevel}, this causes the trajectory to collapse towards the geodesic path. 
The resultant samples fail to exit the artifact-prone regions guarded by the detector boundaries, leading to poor transferability (average ASR drops to 72\%).
Conversely, an excessively large ${\gamma}$ (e.g., $2.0$, see \Cref{fig:simulation}(d)) over-emphasizes artifact avoidance. 
While this strongly pushes the sample away from $\mathbf{z}_1$, it disrupts the delicate semantic alignment with the foundation model. 
This can push the sample into low-density regions of the audio manifold, potentially triggering outlier detection or degrading audio fidelity, which in turn yields diminishing returns in ASR.
In conclusion, the result ${\gamma} \approx 1.0$ confirms that the \textit{Manifold Detour is best achieved when the attractive and repulsive forces are kept in a precise mechanical equilibrium.}

\begin{figure}[t]
    \centering
    \includegraphics[width=0.95\columnwidth]{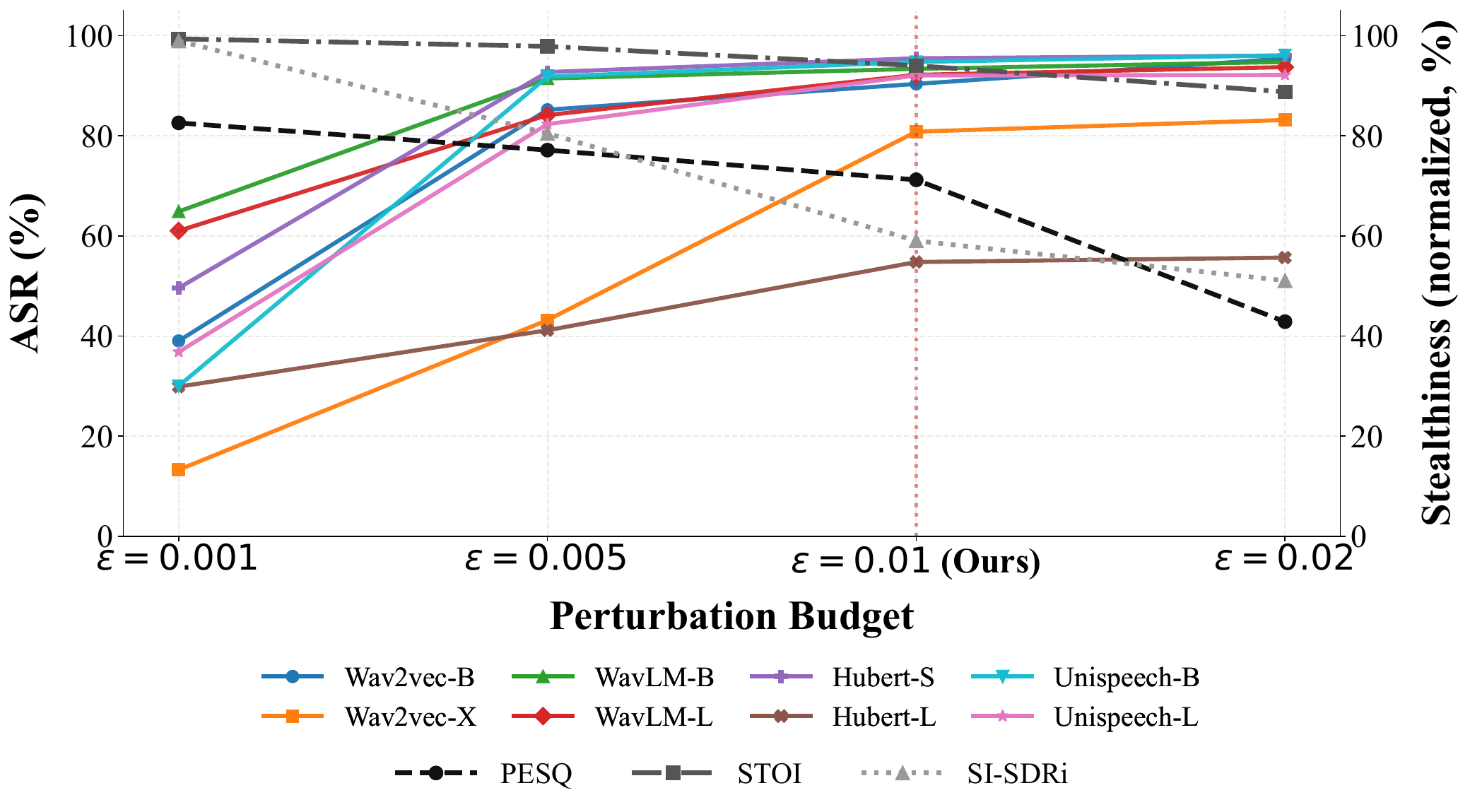}
    % \vspace{-2mm}
    \caption{\small Ablation study on the $\epsilon$ Selection.}
    \label{fig:epsilon_ablation}
\end{figure}

\mypara{Epsilon Ablation}
We further study the effect of the perturbation budget $\epsilon$ on attack effectiveness and stealthiness.
As shown in \Cref{fig:epsilon_ablation}, increasing $\epsilon$ consistently improves the ASR across different SSL backbones, since a larger budget provides a wider feasible space for optimizing adversarial perturbations.
The improvement is especially significant when increasing $\epsilon$ from $0.001$ to $0.005$, where most models exhibit a sharp increase in ASR.
However, after $\epsilon=0.01$, the performance gain becomes marginal: most ASR curves start to saturate, indicating that further enlarging the perturbation budget yields limited additional attack benefit.

Meanwhile, a larger perturbation budget noticeably degrades stealthiness.
When $\epsilon$ increases beyond $0.01$, the perceptual quality metrics drop substantially.
At $\epsilon=0.02$, the adversarial examples obtain PESQ~\cite{rix2001perceptual}, STOI~\cite{taal2011algorithm}, and SI-SDR~\cite{le2019sdr} are $=2.144$, $=0.8878$ and $=20.43$ dB, indicating a clear reduction in audio quality and intelligibility preservation.
Such degradation makes the perturbation less suitable for stealthy attacks, even though the ASR is slightly higher.~\cite{yip2024spgm}

\mypara{Step Ablation and Attack Efficiency}
We further study the impact of the optimization step in the adversarial attack. 
As shown in~\Cref{fig:step_ablation}, reducing the attack budget from Step~30 to Step~20 leads to an average ASR drop of 0.80 percentage points, while Step~5 still remains competitive with an average ASR drop of 1.67 percentage points. 
Increasing the budget from Step~30 to Step~50 brings negligible improvement: the average ASR difference between Step~30 and Step~50 is only 0.02 percentage points, with an average absolute difference of 0.11 percentage points across all surrogate models. 
Since each step takes approximately 0.1 seconds, Step~30 corresponds to only about 3 seconds of optimization per sample, yet achieves nearly the same effectiveness as Step~50. 
These results indicate that our attack reaches high ASR with a small optimization budget, demonstrating strong attack efficiency in addition to effectiveness.

\mypara{Surrogate Model Selection}
\begin{figure}[t]
    \centering
    \includegraphics[width=0.95\columnwidth]{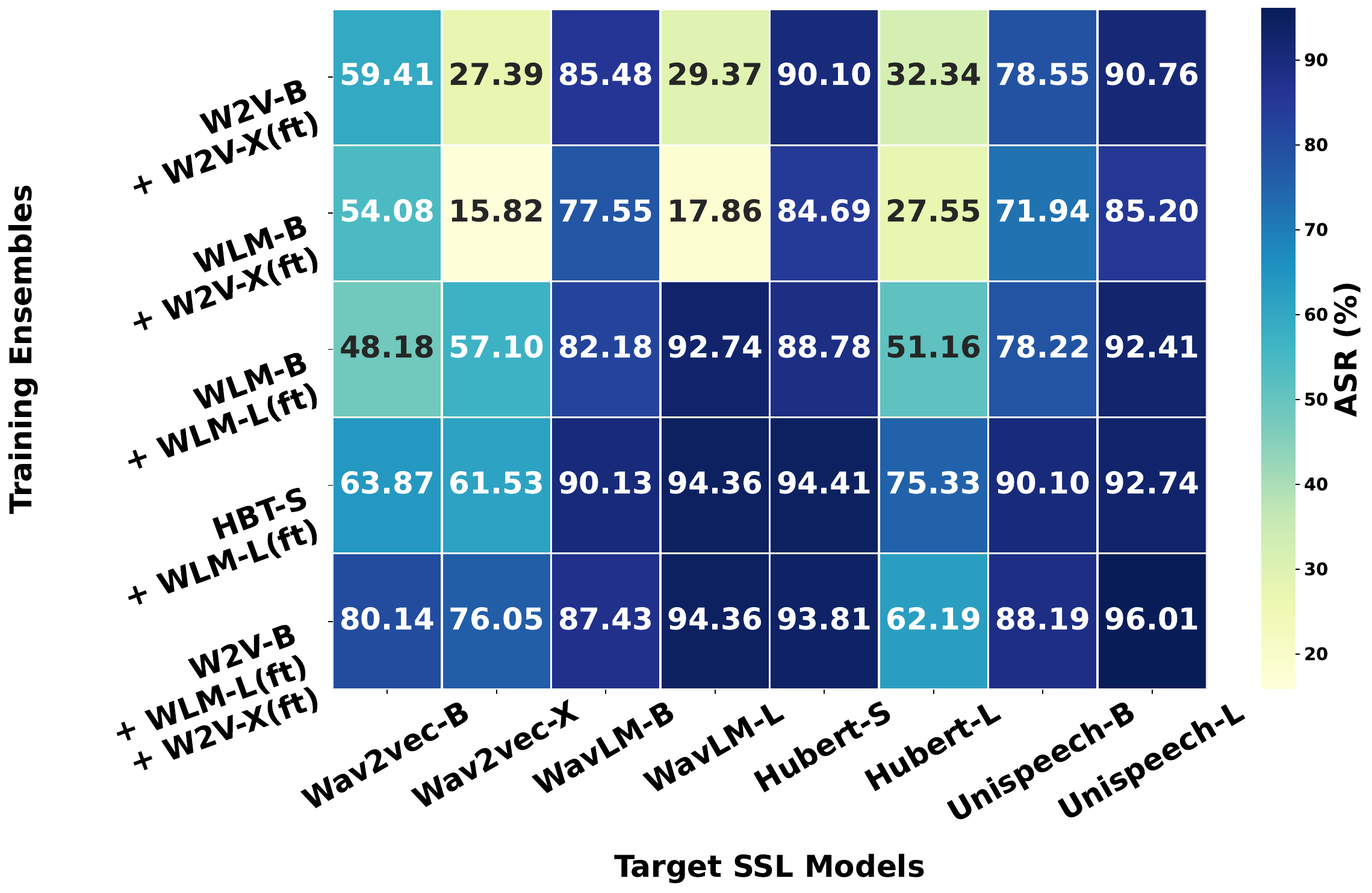}
    % \vspace{-2mm}
    \caption{\small Ablation study on the surrogate model selection.}
    \label{fig:ablation}
\end{figure}
To determine the optimal surrogate configuration, we analyze the transferability across different ensemble settings in \Cref{fig:ablation}. 
We summarize three key guidelines for constructing robust black-box attacks:
\begin{itemize}[leftmargin=*]
    \item \textit{Complementarity over Mere Diversity:}
Homogeneous ensembles (e.g., WavLM Base/Large) suffer from family-specific overfitting, yielding poor transferability to distinct architectures like Wav2vec-X (57.10\%). 
However, architectural differences alone are insufficient. 
The optimal configuration arises from pairing models with \textit{complementary pre-training objectives}: Wav2vec-Base (Contrastive Learning) and fine-tuned WavLM-Large (Masked Prediction). 
This structural heterogeneity creates orthogonal feature constraints, forcing the adversary to bypass the discriminative logic of both paradigms simultaneously.
    \item \textit{The Risk of Destructive Interference:}
Excessive divergence can be detrimental. The pairing of English-centric WavLM-B with multilingual Wav2vec-X results in severe misalignment, leading to \textit{destructive gradient interference} where optimization vectors conflict. 
This suggests that surrogates should share a functional domain while differing in extraction mechanisms.
    \item \textit{Saturation of Ensemble Size:}
While adding a third model yields a numerical peak, the marginal gain is negligible compared to the dual-ensemble. 
This indicates a saturation point: the Wav2vec + WavLM combination already covers the principal directions of the semantic manifold, rendering additional models computationally redundant.
\end{itemize}

\begin{figure}[t]
    \centering
    \includegraphics[width=0.95\linewidth]{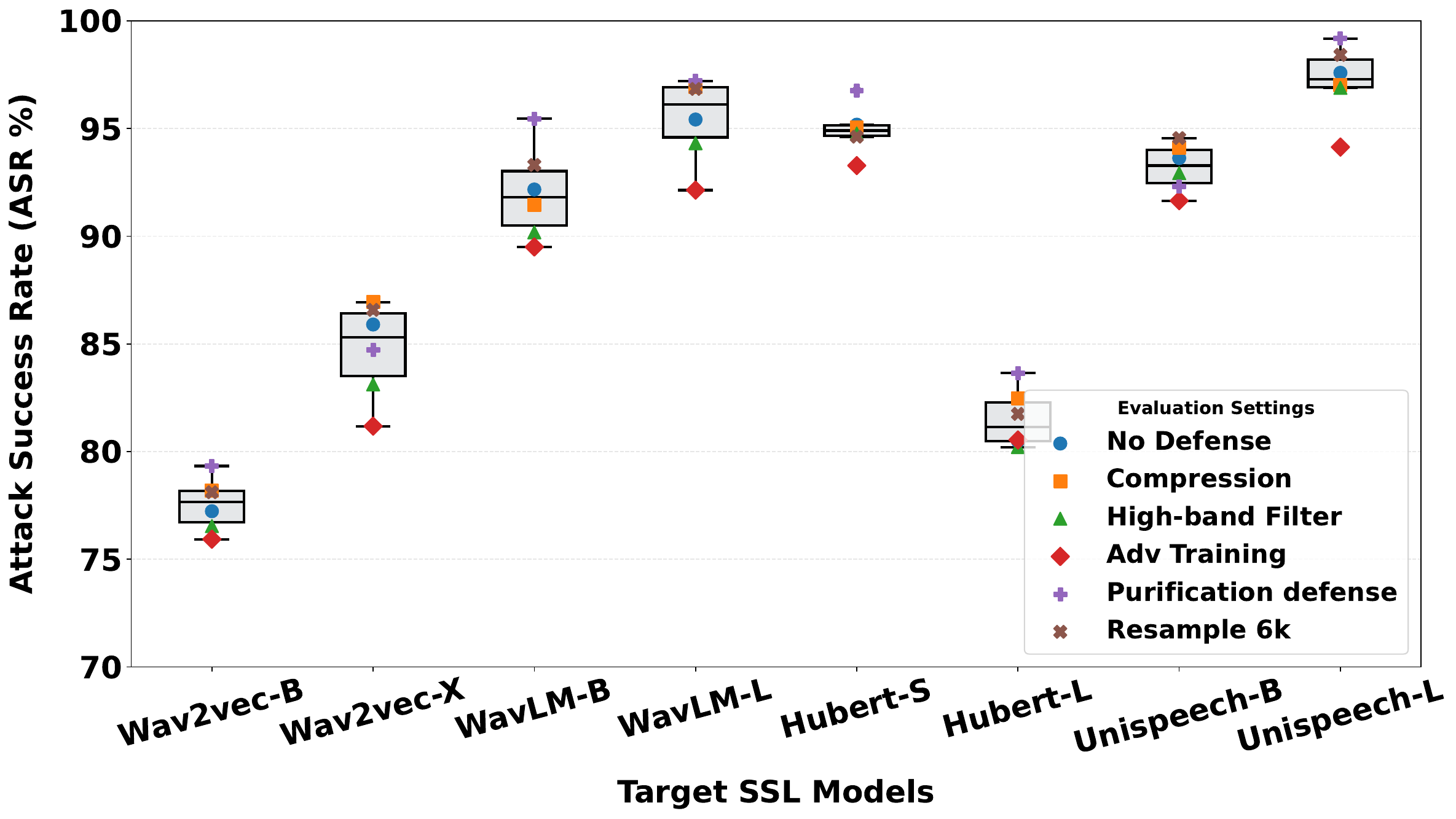}
    % \vspace{-2mm} % 压缩图与标题间距
    \caption{\small Robustness evaluation of our method across eight SSL backbones under various audio defenses in the SLS detector. }
    \label{fig:robustness_line}
    \vspace{-5pt} % 压缩标题与正文间距
\end{figure}

\subsection{Stealthiness and Robustness}
\mypara{Stealthiness Evaluation}
To validate perceptual stealthiness, we conducted a subjective human study with 20 (after filtered) participants evaluating 300 randomly sampled clips. The human study involved 20 volunteers with a background in music to ensure sensitivity to acoustic details. 
The 300 adversarial clips were evaluated in two phases, split by listening environment: 150 tracks via external loudspeakers and 150 tracks via high-fidelity headphones.
In the Single Assessment, participants were asked to flag any suspicious noise without a reference.
In the Comparative Assessment, participants were presented with labeled pairs (Original vs. Adversarial) and asked if they could perceive any difference.  

The study comprised two settings: \textit{Single Assessment} (detecting artifacts in isolation) and \textit{Comparative Assessment} (A/B testing against original audio).
Results confirm high fidelity: 85.25\% of samples were rated as artifact-free in isolation, and 74.75\% were marked as indistinguishable from the original deepfakes in paired comparisons.
We attribute this to our \textit{Dynamic Spectral Mask}, which leverages psychoacoustic masking to confine perturbations strictly to high-energy frequency bands, effectively burying the noise beneath dominant vocal components.

To further verify that MARS preserves the musical and semantic content of singing audio, we evaluate both automatic lyric consistency and standard perceptual quality metrics. We use FireRedASR~\cite{xu2025fireredasr} to transcribe 500 original fake samples and their adversarial counterparts, and compute the word error rate (WER)~\cite{chen2024songbsab} between the two transcriptions. The adversarial samples achieve a same-transcription rate of $85.00\%$, with only three utterances showing transcription differences. Across all evaluated lyrics, the micro WER is only $1.875\%$ and the macro WER is $1.690\%$. The observed differences are minor substitutions or deletions, indicating that the adversarial perturbations rarely alter lyric-level intelligibility. In addition, the adversarial samples maintain high perceptual quality, achieving an average PESQ of $3.56$ ($\pm 0.44$), STOI of $0.94$ ($\pm 0.03$), and SI-SDR of $23.61$ dB ($\pm 1.67$). These results show that MARS can substantially affect detector decisions while preserving both acoustic fidelity and lyric intelligibility. Besides, we provide a side-by-side Mel spectrogram comparison between the original and adversarial audio samples in \Cref{fig:spec_comparison}.

\mypara{Robustness Evaluation}
We evaluate \method under five defense or post-processing settings: MP3 compression to 128kbps, 8KHz high-band filtering, frequency-selective adversarial training (AT)~\cite{zhang2024can}, purification, and 8 kHz resampling.
As shown in \Cref{fig:robustness_line}, \method remains effective across all target SSL models, with ASR staying high under each setting.
AT and high-band filtering cause the most visible drops, but the attack still preserves strong transferability.
Compression, purification, and resampling have little impact and sometimes even increase ASR, suggesting that these transformations may remove detector-relevant synthesis artifacts rather than eliminating the adversarial effect.

\section{Discussion}
\label{sec:discussion}

\mypara{Scope of the Attack}
\method focuses on transferable black-box attacks against SSL-based singing voice deepfake detection (SVDD) systems. This scope is practically relevant because SSL-based detectors currently represent the state-of-the-art paradigm for singing audio deepfake detection, typically using a pre-trained SSL backbone with a task-specific detection head. In addition to standard fine-tuned SSL detectors, we also evaluate variants where the SSL backbone is adapted with LoRA fine-tuning~\cite{hu2022lora} or directly used for detection without fine-tuning~\cite{liu2025nes2net}. In both settings, \method achieves an ASR close to $100\%$, suggesting that the exposed vulnerability is not merely caused by a particular classifier head or fine-tuning strategy, but is rooted in the transferable representation geometry of SSL models. Nevertheless, our study mainly targets the state-of-the-art SSL-based SVDD systems, and does not fully cover future detectors that may incorporate handcrafted acoustic cues, watermark traces or other non-SSL forensic features.

\mypara{Potential Mitigation Strategies}
Our findings suggest that robust SVDD systems should strengthen representation-level robustness instead of only improving the final detection head. 
Potential defenses include adversarial training with diverse SSL backbones and perturbations, consistency checking under common audio transformations, and cross-layer monitoring of abnormal semantic-artifact mismatches. 
In practice, these defenses can be combined with watermark verification, provenance tracking, and ensemble detections.

\section{Conclusion}
This work introduced \method, a meta-adversarial attack for black-box singing voice deepfake detection. 
By combining a surrogate likelihood-based objective with a bi-level tangential optimization strategy, \method achieves strong and transferable attack performance across architectures and domains. 
These findings reveal structural vulnerabilities in SSL-based audio detectors and motivate future defenses beyond linear decision boundaries.

\section*{Ethics Considerations}

This work studies adversarial attacks on singing voice deepfake detection systems with the goal of understanding and improving the robustness of deployed forensic defenses. We recognize that the proposed method is dual-use: while it can expose weaknesses in current detectors, it could also be misused to evade automatic screening systems. To mitigate this risk, our experiments are conducted only on public research datasets and controlled evaluation settings, without targeting any real-world platform, individual singer, or deployed commercial detector. The generated adversarial examples are used solely for robustness analysis and are not intended for unauthorized distribution or misuse. We will follow a responsible release strategy by providing code and artifacts for reproducibility while avoiding content that directly facilitates harmful large-scale evasion. For the human listening study, participants were informed of the study purpose, and all collected responses were anonymized. We hope our findings encourage the development of more robust, transformation-aware, and representation-consistent singing voice deepfake detection systems.

\bibliographystyle{IEEEtran}
\bibliography{reference}

\clearpage
\appendix

\section{Method}
\subsection{Deepfake Effectiveness Performance}

\begin{table}[h]
\centering
\caption{EER \% of different SLS Deepfake Detectors (Mid-layer) on the CtrSVDD dataset.}
\label{tab:ssl_eer_performance}
\renewcommand{\arraystretch}{1.0} % 行高微调，0.9-1.1 之间
\setlength{\tabcolsep}{8pt}       % 列间距微调

% 将这里改成 0.5 到 0.7 之间的数字即可控制大小
% \resizebox{0.3\columnwidth}{!}{% 
\begin{tabular}{@{} l c @{}}
\toprule
\textbf{SSL Models} & \textbf{EER (\%)} \\
\midrule
Wav2vec-B   & 8.39 \\
Wav2vec-X   & 2.62 \\
WavLM-B     & 6.32 \\
WavLM-L     & 4.78 \\
Hubert-S    & 7.13 \\
Hubert-L    & 5.24 \\
Unispeech-B & 9.30 \\
Unispeech-L & 6.55 \\
\bottomrule
\end{tabular}
\end{table}

\subsection{Empirical Performance of Foundation Model's last layer}\label{app:base}

\begin{table}[h]
\centering
\caption{EER \% using only the Last Layer feature of SSL foundation models.}
\label{tab:last_layer_eer}
\footnotesize
\setlength{\tabcolsep}{12pt}
\renewcommand{\arraystretch}{1.1}

\begin{tabular}{@{} l c @{}}
\toprule
\textbf{SSL Models} & \textbf{EER (\%)} \\
\midrule
Wav2vec-B   & 56.17 \\
Wav2vec-X   & 53.38 \\ 
WavLM-B     & 56.39 \\
WavLM-L     & 52.46 \\
Hubert-S    & 54.29 \\
Hubert-L    & 50.72 \\
Unispeech-B & 54.87 \\
Unispeech-L & 51.68 \\
\bottomrule
\end{tabular}
\end{table}

\subsection{Deepfake Effectiveness Performance of different layers in WavLM-large}

\label{app:layer}
To empirically investigate the layer-wise contribution of the SSL backbones to deepfake detection, we randomly sampled 100 singing audio clips and analyzed the top-5 attention layers for each sample during the inference phase. 
Our results reveal a significant concentration of model attention within the intermediate-to-late transformer blocks. 
Specifically, layers 10 through 14 appeared in the top-5 most influential layers with a probability ranging from $62\%$ to $87\%$, substantially exceeding the contribution of other layers. 
This observation suggests that these specific layers encapsulate the most critical acoustic artifacts and discriminative features for deepfake identification. 
A similar trend regarding the importance of these middle layers in audio deepfake tasks has also been reported in~\cite{el2025comprehensive}.

\subsection{Diagnostics for the Anchor-Conditioned Surrogate}
\label{app:anchor_diagnostic}

\mypara{Empirical Support for the Local Evidence Surrogate}
To examine whether the proposed evidence score captures detector-relevant
information, we compare the anchor-based evidence score with the output logits
of the trained MLP detector used in our experiments.
The evidence score achieves a zero-shot AUC of $0.938$ and an EER of $0.113$,
slightly outperforming the trained MLP detector, which obtains an AUC of $0.934$
and an EER of $0.124$.
Moreover, the evidence scores have a Pearson correlation of $r=0.965$ with the
MLP detector logits on a sample-by-sample basis.
These results do not imply that the evidence score is the true victim likelihood
ratio, but they suggest that the local surrogate captures detector-relevant
spoof evidence and provides a useful alternative to directly optimizing a
surrogate classification boundary.

\mypara{Angular-model sanity check}
The vMF assumption is used only as a convenient angular parameterization, not as
a complete generative model of SSL features.
It provides a tractable first-order approximation to the dominant angular
structure of normalized SSL representations.
As a diagnostic, we perform a Kolmogorov-Smirnov goodness-of-fit test on the
fitted one-dimensional marginal of the vMF distribution.
Across both detector and base spaces, the observed marginals are not rejected by
the fitted vMF marginal: in the detector space, the $p$-values are
$p_{\mathrm{real}}=0.68$ and $p_{\mathrm{fake}}=0.59$; in the base space, the
$p$-values are $p_{\mathrm{real}}=0.62$ and $p_{\mathrm{fake}}=0.53$.
These results should be interpreted as a sanity check rather than a proof that
SSL features are exactly vMF-distributed.
Together with the loss-function ablations, they support the use of angular
geometry as a practical approximation for constructing the attack objective.

\subsection{Black-box Separation and Baseline Fairness Protocol}
\label{app:blackbox_baseline_protocol}

\Cref{tab:blackbox_separation} summarizes the separation between the attack-time surrogate models and the victim detectors.
\Cref{tab:baseline_fairness_budget} reports the baseline fairness protocol and effective optimization budgets.
\Cref{tab:hyperparameter_tuning} lists the hyperparameter selection protocol.
All baseline and MARS hyperparameters are fixed before victim evaluation and are selected only using the local surrogate models and validation split.

\section{Additional Spectrogram Visualization}
\label{app:spectrogram_visualization}

\begin{figure*}[htbp]
    \centering
    \begin{subfigure}[b]{0.48\linewidth}
        \centering
        \includegraphics[width=\linewidth]{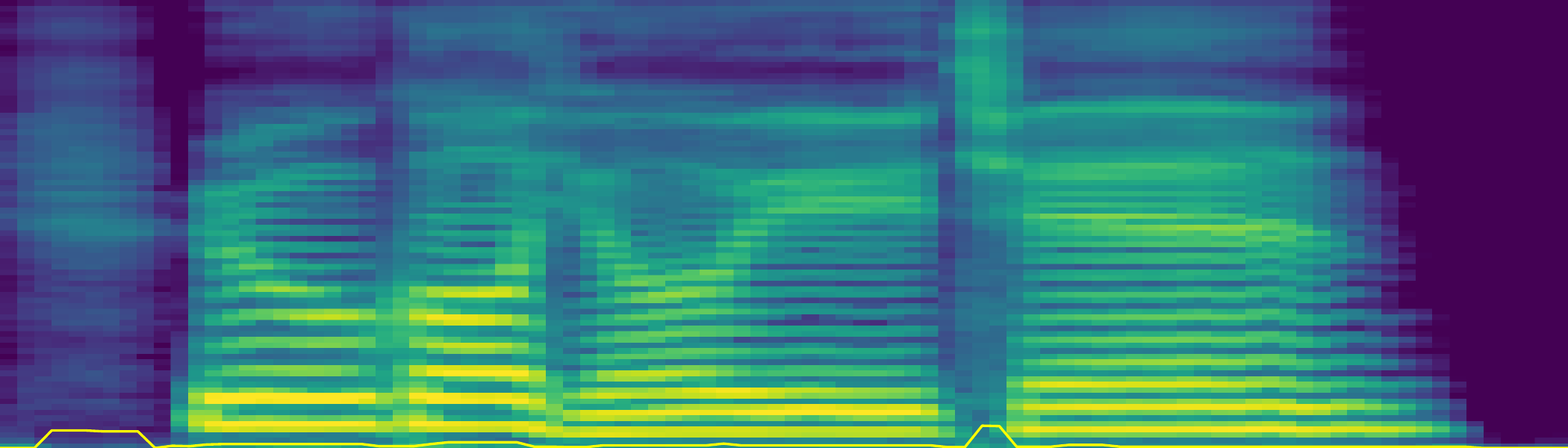}
        \caption{Original Mel Spectrogram}
        \label{fig:spec_real}
    \end{subfigure}
    \hfill
    \begin{subfigure}[b]{0.48\linewidth}
        \centering
        \includegraphics[width=\linewidth]{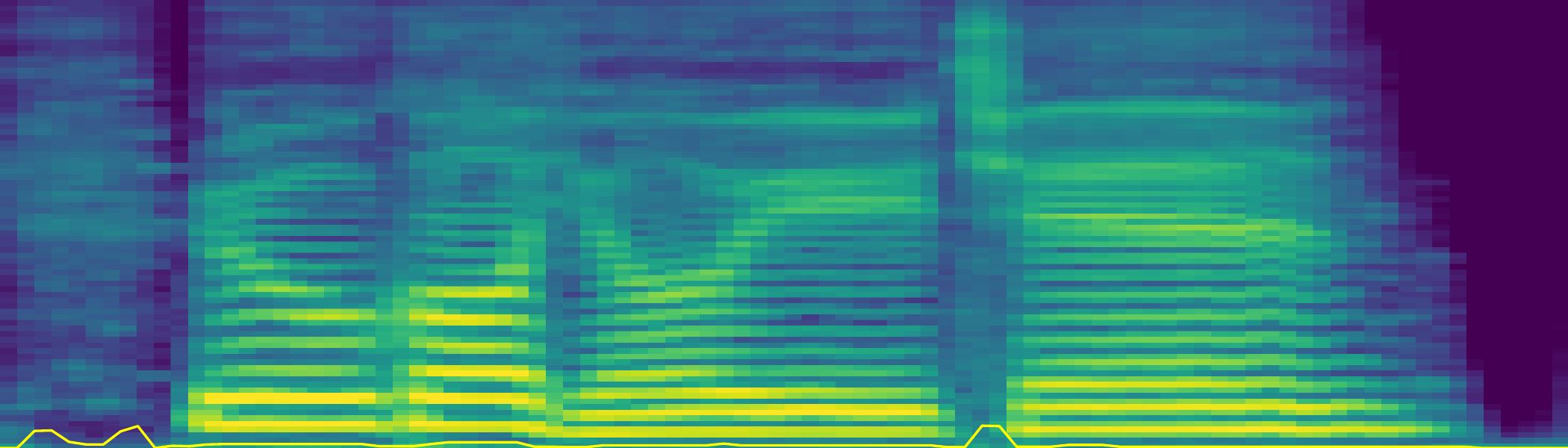}
        \caption{Adversarial Mel Spectrogram}
        \label{fig:spec_fake}
    \end{subfigure}
    
    \caption{Mel spectrogram comparison. (a) The original audio sample. (b) The generated adversarial sample.}
    \label{fig:spec_comparison}
\end{figure*}

\Cref{fig:spec_comparison} presents a side-by-side Mel spectrogram comparison between the original audio sample and the generated adversarial sample.
Visual inspection shows that the adversarial perturbation is highly subtle.
Specifically, the adversarial Mel spectrogram in \Cref{fig:spec_fake} preserves the harmonic structures and formant patterns of the original singing audio in \Cref{fig:spec_real} with high fidelity.
The energy contours, indicated by the yellow curves near the bottom of the spectrograms, also overlap closely between the two samples.
This demonstrates that the proposed attack largely maintains the temporal dynamics and spectral structure of the original utterance, supporting its perceptual stealthiness.

\subsection{Limitation}

While our current evaluation focuses on singing voice deepfake detection models leveraging SSL-based audio representations, this represents only a subset of the broader forensic landscape. Although the geometric principles of our approach are grounded in the general behavior of hyper-spherical embeddings, its efficacy across non-SSL architectures and multi-class detection settings remains to be fully explored. Therefore, we envision extending this framework beyond the audio domain to encompass vision-based tasks, such as image forgery detection and multimodal deepfake analysis. Future investigations will prioritize assessing the cross-modal generalizability of \method to establish it as a more universal adversarial benchmark.

\begin{figure}[t]
    \centering
    \includegraphics[width=0.5\linewidth]{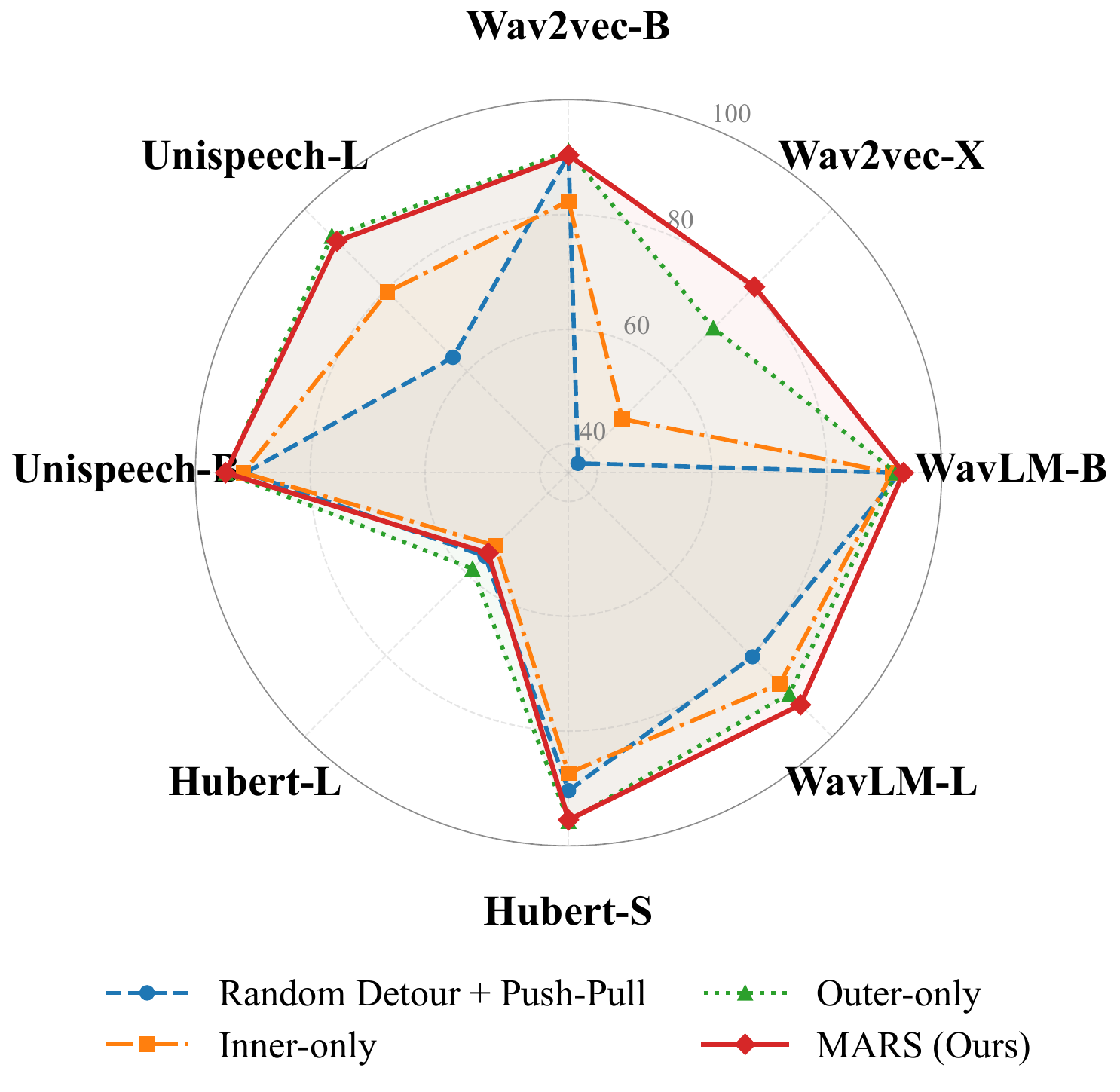}
    \caption{\small Mechanism ablation of the proposed bi-level detour on the MultiConv detector. MARS consistently outperforms random detour, inner-only tangential exploration, and outer-only anchor-conditioned steering across SSL backbones, indicating that the gain comes from the combination of transverse look-ahead and anchor-conditioned evidence steering.}
    \label{fig:detour_rader}
    \vspace{-5pt} 
\end{figure}

\begin{table*}[t]
\centering
\caption{Black-box separation protocol. 
All adversarial examples are generated using local surrogate models only and are transferred to unseen victim detectors without victim queries or victim-specific adaptation.}
\label{tab:blackbox_separation}
\resizebox{\textwidth}{!}{
\begin{tabular}{lccc}
\toprule
\textbf{Component} 
& \textbf{Surrogate used for attack generation} 
& \textbf{Victim used for evaluation} 
& \textbf{Shared with victim?} \\
\midrule
Detection head 
& Lightweight MLP 
& AASIST2 / SLS / MultiConv 
& No \\
Training split 
& CtrSVDD validation split 
& CtrSVDD training split 
& No \\
Detector architecture 
& SSL backbone + MLP surrogate 
& SSL backbone + task-specific SVDD head 
& No \\
Fine-tuning configuration 
& Local surrogate-side fine-tuning only 
& Victim-specific fine-tuning 
& No \\
Victim logits / scores 
& Not observed or used 
& Hidden 
& No \\
Victim labels 
& Not observed or used during attack generation 
& Hidden 
& No \\
Victim thresholds 
& Not observed or used 
& Hidden 
& No \\
Victim gradients / features 
& Not observed or used 
& Hidden 
& No \\
Attack-time victim queries 
& None 
& None 
& No \\
Target-specific adaptation 
& None; one fixed attack configuration is used 
& All victim detectors are evaluated directly 
& No \\
Hyperparameter tuning 
& Surrogate validation only 
& No victim-side tuning 
& No \\
SSL backbone family 
& Wav2vec-B + WavLM-L for MARS 
& Wav2vec / WavLM / HuBERT / UniSpeech 
& Partially; cross-family transfer evaluated \\
\bottomrule
\end{tabular}
}
\end{table*}

\begin{table}[t]
\centering
\caption{Cross-family transfer summary of MARS on CtrSVDD. 
HuBERT and UniSpeech targets are unseen SSL families with respect to the Wav2vec-B / WavLM-L surrogate configuration.}
\label{tab:cross_family_summary}
\resizebox{\linewidth}{!}{
\begin{tabular}{lcc}
\toprule
\textbf{Target SSL family} 
& \textbf{Relation to attack surrogate} 
& \textbf{Average ASR (\%)} \\
\midrule
Wav2vec + WavLM 
& Family-overlap targets 
& 90.38 \\
HuBERT 
& Unseen SSL family 
& 83.12 \\
UniSpeech 
& Unseen SSL family 
& 93.56 \\
HuBERT + UniSpeech 
& Unseen-family targets 
& 88.34 \\
All SSL families 
& All targets 
& 89.36 \\
\bottomrule
\end{tabular}
}
\end{table}

\begin{table*}[t]
\centering
\caption{Baseline fairness and optimization-budget protocol. 
All methods use the same surrogate access, perturbation budget, and evaluation protocol. 
``Grad. evals'' denotes the number of backward gradient evaluations used to generate one adversarial example.}
\label{tab:baseline_fairness_budget}
\resizebox{\textwidth}{!}{
\begin{tabular}{lcccccc}
\toprule
\textbf{Method} 
& \textbf{Implementation} 
& \textbf{Attack objective} 
& \textbf{Surrogate access} 
& \textbf{Constraint} 
& \textbf{Steps} 
& \textbf{Grad. evals} \\
\midrule
PGD 
& Audio waveform 
& Cross-entropy 
& Same MLP surrogate 
& Same $\epsilon$ + spectral mask 
& 60 
& 60 \\
C\&W 
& Audio-domain 
& Margin / confidence loss 
& Same MLP surrogate 
& Same $\epsilon$ + spectral mask 
& 60 
& 60 \\
MI-FGSM 
& Audio-domain 
& Cross-entropy + momentum 
& Same MLP surrogate 
& Same $\epsilon$ + spectral mask 
& 60 
& 60 \\
DI-FGSM 
& Official implementation adapted to waveform 
& Cross-entropy + input diversity 
& Same MLP surrogate 
& Same $\epsilon$ + spectral mask 
& 60 
& 60 \\
TI-FGSM 
& Official implementation adapted to waveform 
& Cross-entropy + translation-invariant smoothing 
& Same MLP surrogate 
& Same $\epsilon$ + spectral mask 
& 60 
& 60 \\
SI-NI-FGSM 
& Official implementation adapted to waveform 
& Cross-entropy + scale invariance + Nesterov momentum 
& Same MLP surrogate 
& Same $\epsilon$ + spectral mask 
& 60 
& 60 \\
VMI-FGSM 
& Official implementation adapted to waveform 
& Cross-entropy + variance tuning 
& Same MLP surrogate 
& Same $\epsilon$ + spectral mask 
& 50 
& 50 + variance samples \\
AWT 
& Official implementation adapted to waveform 
& Cross-entropy + adversarial weight tuning 
& Same MLP surrogate 
& Same $\epsilon$ + spectral mask 
& 50 
& 50 + weight updates \\
Joint Opt. 
& Direct Push-Pull baseline 
& $L_{\rm pull} + \gamma L_{\rm push}$ 
& Same MLP surrogate 
& Same $\epsilon$ + spectral mask 
& 60 
& 60 \\
MARS 
& Proposed 
& Tangential detour + Push-Pull evidence steering 
& Same MLP surrogate 
& Same $\epsilon$ + spectral mask 
& 30 
& 60 \\
\bottomrule
\end{tabular}
}
\end{table*}

\begin{table*}[t]
\centering
\caption{Hyperparameter selection protocol. 
All hyperparameters are selected using only the local surrogate models and validation split before evaluation on victim detectors. 
No victim labels, logits, thresholds, gradients, intermediate features, or evaluation results are used for tuning.}
\label{tab:hyperparameter_tuning}
\resizebox{\textwidth}{!}{
\begin{tabular}{lccc}
\toprule
\textbf{Method} 
& \textbf{Tuned hyperparameters} 
& \textbf{Selected values / protocol} 
& \textbf{Selection criterion} \\
\midrule
PGD 
& Step size 
& $\epsilon=10^{-2}$, step size $=4\times10^{-3}$ 
& Surrogate ASR under quality constraint \\
C\&W 
& Regularization weight, learning rate, confidence margin 
& $c=100$, $\eta=10^{-2}$, $\kappa=0$ 
& Surrogate ASR under quality constraint \\
MI-FGSM 
& Momentum decay, step size 
& $\mu=1.0$, step size $=4\times10^{-3}$ 
& Surrogate ASR under quality constraint \\
DI-FGSM 
& Transformation probability, resize / padding range, step size 
& Official setting adapted to waveform; fixed before victim evaluation 
& Surrogate ASR under quality constraint \\
TI-FGSM 
& Kernel size, smoothing kernel, step size 
& Official setting adapted to waveform; fixed before victim evaluation 
& Surrogate ASR under quality constraint \\
SI-NI-FGSM 
& Number of scales, Nesterov momentum, step size 
& Official setting adapted to waveform; fixed before victim evaluation 
& Surrogate ASR under quality constraint \\
VMI-FGSM 
& Number of variance samples, variance coefficient, step size 
& Official setting adapted to waveform; fixed before victim evaluation 
& Surrogate ASR under quality constraint \\
AWT 
& Weight perturbation scale, update frequency, step size 
& Official setting adapted to waveform; fixed before victim evaluation 
& Surrogate ASR under quality constraint \\
Joint Opt. 
& $\gamma$, step size 
& $\gamma=1.0$, step size $=4\times10^{-3}$ 
& Surrogate ASR under quality constraint \\
MARS 
& $\epsilon$, $\alpha$, $\beta$, $\gamma$ 
& $\epsilon=10^{-2}$, $\alpha=4\times10^{-3}$, $\gamma=1.0$ 
& Surrogate ASR under quality constraint \\
\bottomrule
\end{tabular}
}
\end{table*}

\end{document}